\documentclass[12pt]{iopart}

\usepackage{iopams}  

\usepackage{graphicx}
\usepackage{braket}
\usepackage{amssymb}

\begin{document}
	
	\title[]{Optimal power generation using dark states in dimers strongly coupled to their environment}
	
	\author{D M Rouse$^1$, E M Gauger$^2$ and B W Lovett$^1$}
	\address{$^1$SUPA, School of Physics and Astronomy, University of St. Andrews, St. Andrews KY16 9SS, UK}
	\address{$^2$SUPA, Institute of Photonics and Quantum Sciences, Heriot-Watt University, Edinburgh EH14 4AS, UK}
	\eads{\mailto{dmr9@st-andrews.ac.uk}, \mailto{e.gauger@hw.ac.uk}, \mailto{bwl4@st-andrews.ac.uk}}
	\vspace{10pt}
	\begin{indented}
		\item[]January 2019
	\end{indented}
	
	\begin{abstract}
		Dark state protection has been proposed as a mechanism to increase the power output of light harvesting devices by reducing the rate of radiative recombination. Indeed many theoretical studies have reported increased power outputs in dimer systems which use quantum interference to generate dark states. These models have typically been restricted to particular geometries and to weakly coupled vibrational baths. Here we consider the experimentally-relevant strong vibrational coupling regime with no geometric restrictions on the dimer. We analyze how dark states can be formed in the dimer by numerically minimizing the emission rate of the lowest energy excited eigenstate, and then calculate the power output of the molecules with these dark states. We find that there are two distinct types of dark states depending on whether the monomers form homodimers, where energy splittings and dipole strengths are identical, or heterodimers, where there is some difference. Homodimers, which exploit destructive quantum interference, produce high power outputs but strong phonon couplings and perturbations from ideal geometries are extremely detrimental. Heterodimers, which are closer to the classical picture of a distinct donor and acceptor molecule, produce an intermediate power output that is relatively stable to these changes. The strong vibrational couplings typically found in organic molecules will suppress destructive interference and thus favour the dark-state enhancement offered by heterodimers.
	\end{abstract}
	
	\vspace{2pc}
	
	
	\maketitle

	\section{Introduction}
	
	Organic light harvesting molecules offer the possibility of cheap, stable, flexible and portable photovoltaic devices \cite{ameri2013highly,riede2011efficient}. However, the efficiency of these devices is much lower than those found in more conventional solar cells, with experimentally reported efficiencies reaching up to 13\% after molecular optimization \cite{zhao2017molecular}, but commonly much lower \cite{ameri2013highly,green2017solar}. One way to improve these figures could be to exploit the results of 
	recent theoretical studies, which show that if quantum interference can be harnessed in coupled organic molecules, then the efficiency of organic materials could be much higher~\cite{creatore2013efficient,fruchtman2016photocell,zhang2015delocalized,killoran2015enhancing}.
	
	Pairs of coupled monomers, dimers, are the building blocks of the proposed quantum mechanical light harvesting devices. Modelling has shown that careful optimization of the coupling of the monomers, through tuning their optical properties and relative position and orientation, leads to a hybridization and energy splitting of the single exciton eigenstates. The lower of these states can be optically inactive -- the so-called dark state \cite{creatore2013efficient,fruchtman2016photocell, zhang2016dark, kozyrev2018dark}, whereas the higher lying state has an enhanced transition dipole -- this is the bright state. The dark state is accessed by non-radiative, vibrational relaxation from the bright state. So long as the energy splitting of the bright and dark states is large enough to suppress vibronic re-excitation, then the absorbed energy becomes trapped as exciton recombination is suppressed. This process breaks the detailed balance that leads to the Shockley-Queisser limit, and has been predicted to increase the power output and efficiency of a device built from such molecules~\cite{creatore2013efficient,fruchtman2016photocell,zhang2015delocalized,zhang2016dark,hu2018double,higgins2017quantum,scully2010quantum,dorfman2013photosynthetic,poonia2018solid}. 
	
	Dark state protection has been shown to work effectively for specific geometries of both homodimers~\cite{creatore2013efficient} and heterodimers~\cite{fruchtman2016photocell}, and more recently in linear chains of coupled homodimers~\cite{mattioni2018non}. However, these studies are restricted to the case  of weak environmental coupling; the picture is more complicated in typical organic molecules and dye pigments, since these have strong coupling to their vibrational environments~\cite{mathew2014dye,yang2017rational,sowa2018beyond,burzuri2016sequential}. 
	In this paper, we therefore go beyond these earlier works by treating strong environmental coupling in unrestricted geometries, and including realistic inter-monomer dipole-dipole couplings. We treat the strong coupling by using a polaron transformation~\cite{nazir2016modelling}. In the polaron frame, the monomer energy splittings and inter-monomer coupling become renormalized such that the vibronic environment is in equilibrium with the excited state, rather than displaced by it.
	Energy transfer in the polaron frame has been studied before~\cite{nazir2016modelling,qin2017effects,pollock2013multi}, but not in the context of dark state protection. We will show that strong coupling has a profound effect on quantum interference processes such that it no longer necessarily improves the efficiency of devices. Rather we will propose a different strategy for designing energy harvesting dimer molecules, in which a dark state is chiefly localized on the lower energy monomer.
	
	The paper is structured as follows: Section~\ref{Sec2} begins by introducing the dimer model, before showing how the Hamiltonian is transformed in the polaron frame. In Section~\ref{Sec3} we discuss the Born-Markov master equation formalism before, in Section~\ref{Sec4}, we derive the emission rate from the dark state and explore the conditions under which dark state protection is optimized. We identify different mechanisms for dark state formation for homogeneous and heterogeneous dimers. In Section~\ref{Sec5} we show absorption and emission spectra for the dimers under conditions in which a dark state is expected to form and compare this to conditions which prevent dark state formation. In doing so we provide a way to determine experimentally if a dark state can form. In Section~\ref{Sec6} we explain the theoretical model of power extraction and calculate the power output of dimers over a wide range of realistic parameters. We find that the optimal power output occurs when the dark state is formed in either the homogeneous or heterogeneous dimer and discuss in which regimes either is preferable. Finally, we conclude in Section~\ref{Sec7}.
	
	\section{Model}\label{Sec2} 
\begin{figure}[ht!] 
		\centering
		\includegraphics[width=0.7\linewidth]{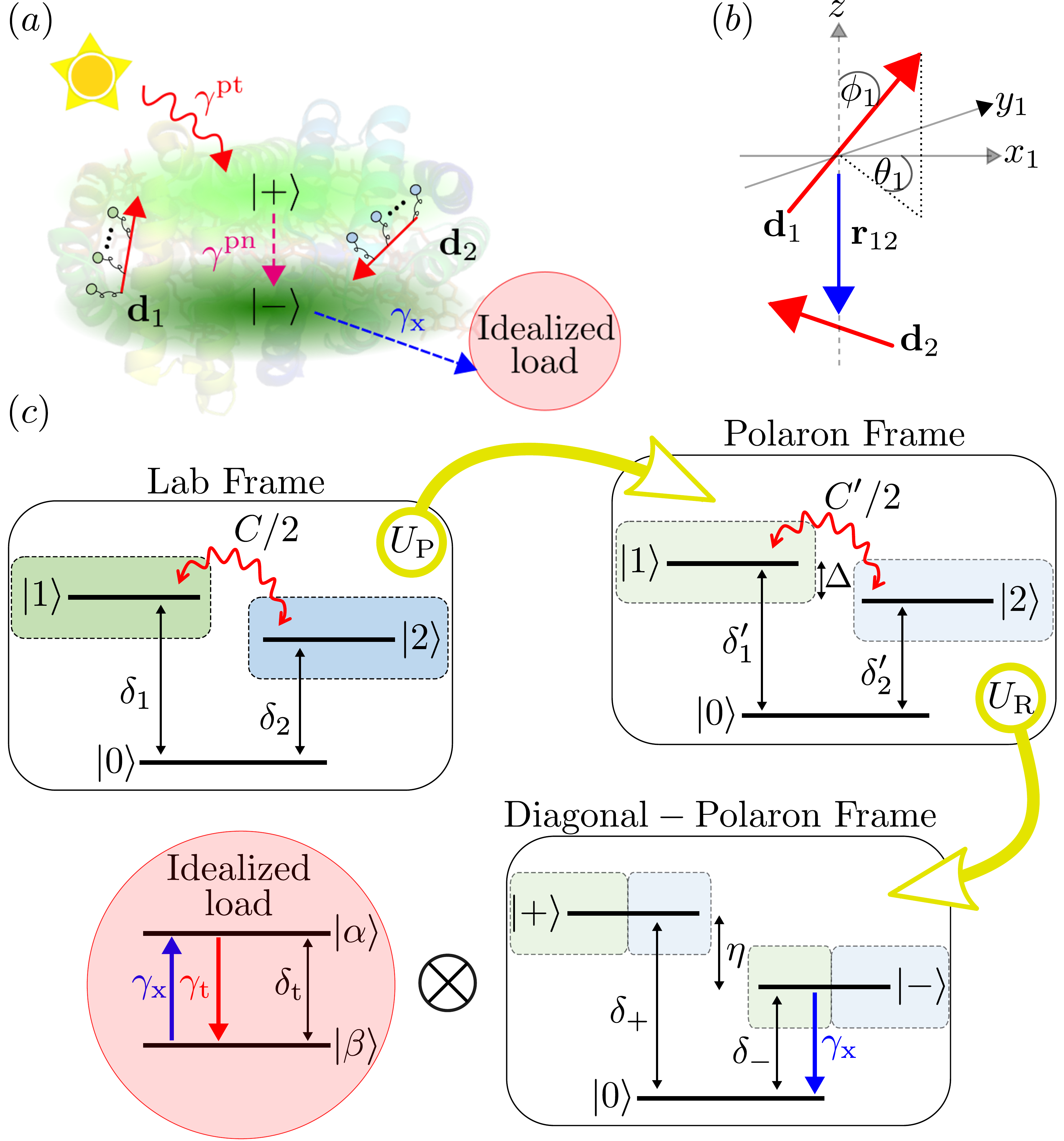} 
		\caption{(a) Cartoon of the energy transfer process involved, from absorption into the dimer to extraction to the idealized load. Monomers are treated as single dipoles with a definite direction and coupling to independent vibrational environments indicated by the masses on springs. (b) Definition of the coordinate system used. (c) Energy level diagrams of the dimer illustrating the system Hamiltonian after each unitary transformation, referred to throughout Section~\ref{Sec2}. The coloured boxes represent that the excited states are coupled to distinct vibrational environments, with the relative filling of each box denoting the coupling strength in that frame. We also show a schematic of the extraction process to the idealized load which will become important when we determine the power output of the dimer in Section~\ref{Sec6}.}
		\label{fig1} 
	\end{figure}

	The dimer consists of two monomers with single dipole moments, $\bi{d}_{j}$ located at positions $\bi{r}_{j}$, where $j=1,2$ labels the dipoles. This system is illustrated in a cartoon in Figure~\ref{fig1}(a) where the dipoles exist as part of a larger protein structure. By virtue of the dipole-dipole coupling, the monomers form symmetric and antisymmetric eigenstates (denoted $\ket{+}$ and $\ket{-}$ respectively), depicted as green clouds. In an ideal scenario quantum interference is constructive at the $\ket{+}$, leading to enhanced photon absorption, and is conversely suppressed at the $\ket{-}$, forming a dark state. After absorption at the brighter $\ket{+}$, the excitation is transferred non-radiatively to the darker $\ket{-}$ where it becomes trapped until extraction to an idealized load to produce power. We will now describe the dimer system mathematically.
	\subsection{Dimer Hamiltonian}
	We assume that each monomer can be treated as a two level system. The excited states are denoted $\ket{1}$ and $\ket{2}$, and the ground state as $\ket{0}$. This lab frame energy level diagram is shown in Figure~\ref{fig1}(c). We use natural units throughout ($\hbar=c\equiv1$). The Hamiltonian describing this system part is 
	\begin{equation}
	H_{\mathrm{S}}=\sum_{j=1}^2\delta_{j}\ket{j}\bra{j}+\frac{C}{2}\left(\ket{1}\bra{2}+\ket{2}\bra{1} \right),
	\label{eq:HamSystem}
	\end{equation}
	where $\delta_j$ is the energy splitting of monomer-$j$ and $C$ is the static dipole-dipole coupling:
	\begin{equation}
	\frac{C}{2}=\frac{1}{4\pi r_{12}^3}\left[\bi{d}_1 \cdot \bi{d}_2 -3\left(\bi{d}_1 \cdot \bi{r}_{12}\right)\left(\bi{d}_2 \cdot \bi{r}_{12}\right)\right],
	\label{eq:DipoleDipoleCoupling}
	\end{equation}  
	where $\bi{r}_{12}=\bi{r}_1-\bi{r}_2$ is the separation vector of the monomers. No restrictions are placed on any of the system parameters. We only consider the single excitation subspace since under illumination with solar temperature radiation the dimer will very rarely hold two excitations.
	
	Each monomer couples to a global, multimode photon environment described by $
	H_{\mathrm{E}}^{\gamma}=\sum_{\bi{q},\lambda}\nu_{\bi{q}}a^{\dagger}_{\bi{q},\lambda}a_{\bi{q},\lambda}$ where $\nu_{\bi{q}}$ is the frequency of mode $\bi{q}$ and $a^{\dagger}_{\bi{q},\lambda}$ ($a_{\bi{q},\lambda}$) is the creation (annihilation) operator for a photon of mode $\bi{q}$ and polarization $\lambda$. Under the rotating wave approximation, the interaction of monomer-$j$ with this field results in the Hamiltonian term,
	\begin{equation}
	H_{\mathrm{I}}^{\gamma,j} = \rmi \ket{0}\bra{j} \sum_{\bi{q}, \lambda} \bi{d}_{j} \cdot \bi{u}_{\bi{q}\lambda}^*(\bi{r}_{j})a_{\bi{q}\lambda}^{\dagger}+\mathrm{H.c.},
	\label{eq:HamPhotonInteraction}
	\end{equation}
	where H.c. denotes the Hermitian conjugate.\footnote{We have directly verified that neglecting counter rotating terms in (\ref{eq:HamPhotonInteraction}), and ignoring the double excited state, makes negligible difference to any results we present.} We take the spatial mode functions $u_{\bi{q}\lambda}(\bi{r})$ of the field to be those of free space with volume $V$ and permittivity $\epsilon$,
	\begin{equation}
	u_{\bi{q}\lambda}(\bi{r})=\sqrt{\frac{\nu_{\bi{q}}}{2\epsilon V}}\bi{e}_{\bi{q}\lambda}e^{i\bi{q}\cdot\bi{r}},
	\label{eq:SpatialModeFunction}
	\end{equation}
	where $\bi{e}_{\bi{q}\lambda}$ are unit vectors describing the polarization state.
	
	Finally, the monomers interact with their own local phonon environment $H_{\mathrm{E}}^{\mathrm{pn},j} = \sum_{\bi{k}}\omega_{\bi{k},j}b_{\bi{k},j}^{\dagger}b_{\bi{k},j}$ where $\omega_{\bi{k},j}$ is the frequency of a phonon of mode $\bi{k}$ within monomer-$j$ and $b^{\dagger}_{\bi{k},j}$ ($b_{\bi{k},j}$) is the creation (annihilation) operator for that phonon. These interactions are represented by the usual displacement of the excited states,
	\begin{equation}
	H_{\mathrm{I}}^{\mathrm{pn},j} = \ket{j}\bra{j}\sum_{\bi{k}}(g_{\bi{k},j}b_{\bi{k},j}^{\dagger} + g_{\bi{k},j}^*b_{\bi{k},j}),
	\label{eq:HamPhononInt}
	\end{equation}
	with coupling strength $g_{\bi{k},j}$. The full Hamiltonian is then \begin{equation}
	H=H_{\mathrm{S}}+\sum_{j=1}^2(H_{\mathrm{I}}^{\gamma,j}+H_{\mathrm{I}}^{\mathrm{pn},j}+H_{\mathrm{E}}^{\mathrm{pn},j})+H_{\mathrm{E}}^{\gamma}.
	\label{eq:HamFull}
	\end{equation}
	\subsection{Polaron transformation}\label{Sec2.2}
	\label{Sec:PolaronTransformation}
	The presence of large Stokes' shifts in candidate monomers for implementing dark state protection ideas, such as those found in~\cite{fruchtman2016photocell}, indicates that these molecules have strong coupling to vibrational modes. Therefore, we transform to the polaron frame which takes some of the phonon interaction Hamiltonian into the system, leaving a residual interaction that can be treated within a weak coupling theory. 
	
	The polaron transformation is generated by the unitary operator $U_{\mathrm{P}}=\rme^G$ where $G=\sum_{\bi{k},j}\ket{j}\bra{j}(g_{\bi{k},j}b^{\dagger}_{\bi{k},j}-g^*_{\bi{k},j}b_{\bi{k},j})/\omega_{\bi{k},j}$. This can be decomposed into the dipole basis as $\rme^{\pm G}=\ket{0}\bra{0}+\sum_{j} B^{\pm}_{j}\ket{j}\bra{j}$, where
	\begin{equation}
	B^{\pm}_{j}=\mathrm{exp}[\pm\sum_{\bi{k}}(g_{\bi{k},j}b^{\dagger}_{\bi{k},j}-g^*_{\bi{k},j}b_{\bi{k},j})/\omega_{\bi{k},j}]\equiv \prod_{\bi{k}}D_{\bi{k},j}\left(\pm \frac{g_{\bi{k},j}}{\omega_{\bi{k},j}}\right),
	\label{eq:DisplaceOper}
	\end{equation}
	is a product of displacement operators, themselves defined by $D_{\mathrm{x}}(\alpha)=\rme^{\alpha b^{\dagger}_{\mathrm{x}}-\alpha^*b_{\mathrm{x}}}$~\cite{nazir2016modelling,qin2017effects,pollock2013multi}.

	After applying the polaron transformation to the full Hamiltonian~(\ref{eq:HamFull}), we subsequently partition the terms into system, environment and interaction Hamiltonians in the usual way~\cite{nazir2016modelling,pollock2013multi,scerri2017method}. We find that both the photon and phonon environment Hamiltonians are unchanged. The system Hamiltonian becomes
	\begin{equation}
	H_{\mathrm{SP}}=\sum_{j=1}^2\delta_{j}^{\prime}\ket{j}\bra{j}+\frac{C^{\prime}}{2}\left(\ket{1}\bra{2}+\ket{2}\bra{1} \right),
	\label{eq:HamSystemPolaron}
	\end{equation}
	where the primed variables indicate renormalization by phonon interactions. Specifically, defining the spectral density of the phonon environment coupled to monomer-$j$ as
	\begin{equation}
	J_{j}(\omega)=\sum_{\bi{k}}\left|{g_{\bi{k},j}}\right|^2\delta(\omega-\omega_{\bi{k},j}),\label{eq:SpectralDensity}
	\end{equation}
	we have $\delta_{j}^{\prime}=\delta_{j}-\lambda_{j}$ where
	\begin{equation}
	\lambda_j=\int_0^{\infty}\rmd \omega\frac{ J_{j}(\omega)}{\omega},
	\label{eq:renorm}
	\end{equation}
	is the reorganization energy of the phonon environment, and $C^{\prime}=\kappa_1\kappa_2C$ where
	\begin{equation} 
	\kappa_{j}\equiv\langle B_{j}^{\pm}\rangle_{\mathrm{pnj}}=\rme^{-\frac{1}{2}\phi_{j}(0)},
	\label{eq:kappa}
	\end{equation}
	and 
	\begin{equation}
	\phi_{j}(t)=\int_0^{\infty}\rmd \omega\frac{J_{j}(\omega)}{\omega^2}\left[\cos(\omega t)\coth\left(\frac{\beta_{\mathrm{pn}}\omega}{2}\right)-{\rmi}\sin(\omega t)\right],\label{eq:PhononProp}
	\end{equation}
	is the phonon propagator for the phonon environment with inverse temperature $\beta_{\mathrm{pn}}~=~\left(k_{\mathrm{B}}T_{\mathrm{pn}}\right)^{-1}$. The polaron-frame system Hamiltonian is depicted in Figure~\ref{fig1}(c). In the polaron frame, the photon and phonon interaction Hamiltonians become
	\begin{eqnarray}
	H_{\mathrm{IP}}^{\gamma}=\sum_{j=1}^2H_{\mathrm{IP}}^{\gamma,j} = \sum_{j=1}^2\rmi \ket{0}\bra{j} B_{j}^- \sum_{\bi{q}, \lambda} \bi{d}_{j} \cdot \bi{u}_{\bi{q}\lambda}^*(\bi{r}_{j})a_{\bi{q}\lambda}^{\dagger}+\mathrm{H.c.},\\
	H_{\mathrm{IP}}^{\mathrm{pn}}=\frac{C}{2}\left(\mathcal{B}\ket{1}\bra{2}+\mathcal{B}^{\dagger}\ket{2}\bra{1}\right),
	\label{eq:HamPhotonInteractionPolaron}
	\end{eqnarray}
	where $\mathcal{B}=B_1^+B_2^--\kappa_1\kappa_2$, and note that $(B_{j}^+)^{\dagger}=B_{j}^-$.

	\subsection{Diagonalization}\label{Sec2.3}
	The stronger the dipole-dipole coupling $C^{\prime}$ is compared to the detuning of the monomers $\Delta$ in the polaron frame, the more delocalized the excitons are over the monomers~\cite{kringle2018temperature}. The phase relationship between the monomer components of the eigenstate wave functions then determines their optical dipole matrix element with the ground state, and fine tuning the degree of delocalization leads to the formation of dark states for various monomers~\cite{fruchtman2016photocell}.
	
	The diagonalized system Hamiltonian (\ref{eq:HamSystemPolaron}) is written $\tilde{H}_{\mathrm{SP}}=\sum_{\sigma=\pm}\delta_{\sigma}\ket{\sigma}\bra{\sigma}$, where the eigenvalues are
	\begin{equation}
	\delta_{\pm}=\frac{1}{2}\left(\delta_1^{\prime}+\delta_2^{\prime} \pm \eta \right),
	\end{equation}
	with eigenstate detuning $\eta=\sqrt{\Delta^2+C^{\prime 2}}$, and renormalized monomer detuning $\Delta~=~\delta_1^{\prime}-\delta_2^{\prime}$. The symmetric and antisymmetric eigenstates (see Figure~\ref{fig1}(c) for a depiction) are
	\numparts
	\begin{eqnarray}
	\ket{+}=\cos\frac{\chi}{2}\ket{1}+\sin\frac{\chi}{2}\ket{2},\\
	\ket{-}=-\sin\frac{\chi}{2}\ket{1}+\cos\frac{\chi}{2}\ket{2}\label{eq:MinusES},
	\end{eqnarray}
	\endnumparts
	where
	\numparts
	\begin{eqnarray}
	\cos\chi=\frac{\Delta}{\eta},\label{eq:chi1}\\
	\sin\chi=\frac{C^{\prime}}{\eta}.\label{eq:chi2}
	\end{eqnarray}
	\endnumparts
	Although we have named the $\ket{+}$ and the $\ket{-}$ the symmetric and antisymmetric eigenstates respectively, this definition is strictly only correct for positive couplings, $C$. When the coupling becomes negative, the symmetry of the eigenstates swap. In the eigenbasis, the photon interaction Hamiltonian becomes
	\begin{equation}
	\tilde{H}_{\mathrm{IP}}^{\gamma} = \sum_{\sigma=\pm}\rmi \ket{0}\bra{\sigma} \sum_{\bi{q}, \lambda} \big(D^{(\sigma)}_{\bi{q}\lambda}(\bi{r}_{1},\bi{r}_{2})\big)^{\dagger}a_{\bi{q}\lambda}^{\dagger}+\mathrm{H.c.},
	\label{eq:PhotonPolaronInt}
	\end{equation}
	where
	\numparts
	\begin{eqnarray}
	D^{(+)}_{\bi{q}\lambda}(\bi{r}_1,\bi{r}_2)=\cos\frac{\chi}{2}\bi{d}_1\cdot\bi{u}_{\bi{q}\lambda}(\bi{r}_1)B^{+}_1+\sin\frac{\chi}{2}\bi{d}_2\cdot\bi{u}_{\bi{q}\lambda}(\bi{r}_2)B^{+}_2,\label{eq:EigenCoupling1}\\
	D^{(-)}_{\bi{q}\lambda}(\bi{r}_1,\bi{r}_2)=-\sin\frac{\chi}{2}\bi{d}_1\cdot\bi{u}_{\bi{q}\lambda}(\bi{r}_1)B^{+}_1+\cos\frac{\chi}{2}\bi{d}_2\cdot\bi{u}_{\bi{q}\lambda}(\bi{r}_2)B^{+}_2.\label{eq:EigenCoupling2}
	\end{eqnarray}
	\endnumparts
	Finally, the phonon Hamiltonian becomes
	\begin{equation}
	\tilde{H}_{\mathrm{IP}}^{\mathrm{pn}}=\sum_{\alpha=x,y,z}\tau_{\alpha}B_{\alpha},
	\end{equation}
	where we have introduced the Pauli operators
	\numparts
	\begin{eqnarray}
	\tau_{\mathrm{x}}=\ket{+}\bra{-}+\ket{-}\bra{+},\\ \tau_{\mathrm{y}}=-\rmi\left(\ket{+}\bra{-}-\ket{-}\bra{+}\right),\\ \tau_{\mathrm{z}}=\ket{+}\bra{+}-\ket{-}\bra{-},
	\end{eqnarray}
	\endnumparts
	and the following groupings of phonon operators
	\numparts
	\begin{eqnarray}
	B_{\mathrm{x}}=\frac{C^{\prime}}{2}(\mathcal{B}^{\dagger}+\mathcal{B})\cos\chi,\label{eq:PhononOpx}\\ B_{\mathrm{y}}=\frac{C^{\prime}}{2\rmi}(\mathcal{B}^{\dagger}-\mathcal{B}),\label{eq:PhononOpy}\\
	B_{\mathrm{z}}=\frac{C^{\prime}}{2}(\mathcal{B}^{\dagger}+\mathcal{B})\sin\chi.\label{eq:PhononOpz}
	\end{eqnarray}
	\endnumparts
	Thus, the full Hamiltonian in the polaron frame reads:
	\begin{equation}
	\tilde{H}_{\mathrm{P}}=\tilde{H}_{\mathrm{SP}}+\tilde{H}^{\gamma}_{\mathrm{IP}}+\tilde{H}^{\mathrm{pn}}_{\mathrm{IP}}+H^{\gamma}_{\mathrm{E}}+\sum_{j=1}^2H^{\mathrm{pn},j}_{\mathrm{E}}.
	\end{equation}

	\section{Master equation formalism}\label{Sec3}
	
	To calculate the energy transfer dynamics we use the standard second order Born-Markov master equation in the polaron frame. A full discussion on the master equation formalism can be found in~\cite{breuer2002theory}. The Born-Markov assumptions are that the density matrix $\rho(t)$ is separable at all times into a system and environment part, and that the environment retains no memory of its interactions with the system. Therefore we can write the total density matrix in the factorized form $\rho(t)=\rho_{\mathrm{S}}(t)\otimes\rho_{\mathrm{E}}^{\mathrm{pn}}\otimes\rho_{\mathrm{E}}^{\gamma}$ where $\rho_{\mathrm{S}}(t)$ is the system density matrix, and $\rho_{\mathrm{E}}^x$ are the density matrices for boson environments $x$, and are constant in time and in thermal equilibrium, i.e.
	\begin{equation}
	\rho_{\mathrm{E}}^x=\frac{\rme^{-\beta_xH_{\mathrm{E}}^x}}{\tr_{\mathrm{E},x}\left(\rme^{-\beta_xH_{\mathrm{E}}^x }\right)},
	\end{equation}
	where $\beta_x$ ($\tr_{\mathrm{E},x}$) is the inverse temperature of (trace over) environment $x$. We are interested in the dynamics of the system, and because of the density matrix partitioning we are able to trace out the environment degrees of freedom and derive a dynamical equation  $\rho_{\mathrm{S}}(t)=\tr_{\mathrm{E}}\left[\rho(t)\right]$ where E denotes both environments.
	
	The closed evolution of the system is then treated exactly, but the effects of the system-environment interactions on the dynamics are treated perturbatively to second order within the Born-Markov approximation. To summarize a standard result, one finds that in the Schr\"{o}dinger picture,
	\begin{equation}
	\frac{\rmd}{\rmd t}\rho_{\mathrm{S}}(t)=-\rmi[\tilde{H}_{\mathrm{SP}},\rho_{\mathrm{S}}(t)]+\mathcal{D}^{\gamma}\left(\rho_{\mathrm{S}}(t)\right)+\mathcal{D}^{\mathrm{pn}}\left(\rho_{\mathrm{S}}(t)\right),
	\label{eq:SysSchrotoint}
	\end{equation}
	where $\mathcal{D}^x(\rho)$ are the dissipators for environment $x$. After writing the interaction Hamiltonians as $\tilde{H}_{\mathrm{IP}}^x=\sum_{\alpha}A_{\alpha}^x\otimes C_{\alpha}^x$ where the $A_{\alpha}^x$ and $C_{\alpha}^x$ operators are in the system and environment(s) Hilbert spaces respectively, then in the Schr\"odinger picture (and polaron frame) the dissipators have the form
	\begin{eqnarray}
	\mathcal{D}^x\left(\rho_{\mathrm{S}}(t)\right)=\sum_{\alpha,\beta}\sum_{\omega,\omega^{\prime}}\Xi^x_{\alpha\beta}(\omega)\left[A_{\beta}^x(\omega)\rho_{\mathrm{S}}(t)A_{\alpha}^{x\dagger}(\omega^{\prime})-A_{\alpha}^{x\dagger}(\omega^{\prime})A_{\beta}^x(\omega)\rho_{\mathrm{S}}(t) \right] \nonumber\\
	+ \mathrm{H.c.},
	\label{eq:Dissipator}
	\end{eqnarray}
	where the rates of each process are given by the environment correlation functions (ECFs)
	\begin{equation}
	\Xi^x_{\alpha\beta}(\omega)=\int_0^{\infty}\rmd t\rme^{\rmi\omega t}\langle C_{\alpha}^{x\dagger}(t)C_{\beta}^x(0) \rangle_{\mathrm{E}}.
	\label{eq:ECF}
	\end{equation}
	In (\ref{eq:Dissipator}), the system operators are in eigenoperator form, defined as $A_{\alpha}^x(\omega)=\sum_{\epsilon}\mathcal{P}(\epsilon)A_{\alpha}^x\mathcal{P}(\epsilon+\omega)$ where $\mathcal{P}(\epsilon)=\ket{\epsilon}\bra{\epsilon}$, $\tilde{H}_{\mathrm{SP}}\ket{\epsilon}=\epsilon\ket{\epsilon}$ and the sum extends over all eigenvalues $\epsilon$. Similarly, the environment operators in (\ref{eq:ECF}) are evaluated in the interaction picture.
	
	We summarize the contributions to the master equation from each term of (\ref{eq:SysSchrotoint}) in \ref{App:ME}. In the following we only explicitly discuss terms which influence the formation of the dark state.

	\section{Creating dark states}\label{Sec4}
	For an eigenstate to be dark, the emission of a photon must be suppressed. It is also necessary that the dominant role of phonon processes is to cause excitations to be transferred into the dark state from the bright state, rather than vice-versa. The latter is easily achieved by choosing our target dark state as the lower energy, antisymmetric eigenstate, $\ket{-}$. The former is only achieved by carefully tuning all parameters of the system, and the phonon coupling strength, such that the photon emission rate is suppressed from the $\ket{-}$. 
	
	\subsection{Antisymmetric eigenstate photon rates.}
	We derive the photon emission and absorption rates in \ref{App:DarkEmission} but here we quote the results. We find that the rates between the $\ket{-}$ and the ground state can be written as
	\begin{equation}
	\gamma_-^{\mu}=\sin^2\frac{\chi}{2}\Gamma_1^{\mu}(\delta_-)+\cos^2\frac{\chi}{2}\Gamma_2^{\mu}(\delta_-)-\sin\chi\Gamma_{12}^{\mu}(\delta_-),
	\label{eq:InitialMinusPhotonRate}
	\end{equation}
	where $\mu=\mathrm{A},\mathrm{E}$ for absorption and emission, and the angle $\chi$ is defined by (\ref{eq:chi1}) and (\ref{eq:chi2}). In (\ref{eq:InitialMinusPhotonRate}), $\Gamma_j^{\mu}(\omega)$ are the rates for the phonon-renormalized electronic transitions of monomer-$j$ and $\Gamma_{12}^{\mu}(\omega)$ are the contributions to the rates due to the collective effects of the monomers in the dimer. Therefore, recalling the definition of the $\ket{-}$ (\ref{eq:MinusES}), the first two terms in (\ref{eq:InitialMinusPhotonRate}) are seen to describe the contribution from each monomer independently. These include a trigonometric factor describing the localization on either monomer, as well as the electronic transition rate of the associated monomer. The final term describes the effect of the coupling between the monomers and the overall sign determines if this leads to constructive ($>0$) or destructive ($<0$) interference. A term due to destructive interference in the rate expression is beneficial for creating a dark state.
	
	The terms due to the monomer coupling, $\Gamma^{\mu}_{12}(\omega)$ can be derived analytically (see \ref{App:DarkCollective}) and for absorption and emission are found to be
	\numparts
	\begin{eqnarray}
	\Gamma_{12}^{\mathrm{A}}(\omega)=\kappa_1\kappa_2\sqrt{\gamma_1(\omega)\gamma_2(\omega)}\mathcal{F}(\omega r_{12})N(\omega),\\
	\Gamma_{12}^{\mathrm{E}}(\omega)=\kappa_1\kappa_2\sqrt{\gamma_1(\omega)\gamma_2(\omega)}\mathcal{F}(\omega r_{12})\left[1+N(\omega)\right],
	\end{eqnarray}
	\endnumparts
	where 
	\begin{equation}
	\gamma_j(\omega)=(d_j^2\omega^3)/(3\pi),
	\label{eq:BETR}
	\end{equation} 
	is the bare electronic transition rate, $N(\omega)=\left(\rme^{\beta_{\gamma}\omega}-1\right)^{-1}$ is the photon Bose-Einstein distribution and $\kappa_j$ are defined by (\ref{eq:kappa}). The cross function is
	\begin{equation}
	\mathcal{F}(x)=\frac{3}{2}\left(\alpha_{12}\frac{\sin x}{x}+\beta_{12}\left[\frac{\cos x}{x^2}-\frac{\sin x}{x^3}\right]\right),
	\label{eq:FFunction}
	\end{equation}
	with
	\begin{eqnarray}
	\alpha_{12}=\hat{\bi{d}}_1\cdot\hat{\bi{d}}_2-(\hat{\bi{d}}_1\cdot\hat{\bi{r}}_{12})(\hat{\bi{d}}_2\cdot\hat{\bi{r}}_{12}),\\
	\beta_{12}=\hat{\bi{d}}_1\cdot\hat{\bi{d}}_2-3(\hat{\bi{d}}_1\cdot\hat{\bi{r}}_{12})(\hat{\bi{d}}_2\cdot\hat{\bi{r}}_{12}),
	\end{eqnarray}
	where the hat denotes the unit vector. The range of the cross function is $-{1\le\mathcal{F}(\omega r_{12})\le1}$ and its value indicates the effect of the delocalization on the photon rates (\ref{eq:InitialMinusPhotonRate}). In realistic dimer systems, the eigenstate energy splittings $\delta_{\pm}\sim\mathrm{eV}$ and monomer separation $r_{12}\sim\mathrm{nm}$, in which case we can use the following limit
	\begin{equation}
	\mathcal{F}\equiv\lim_{\omega r_{12}\to 0} \mathcal{F}(\omega r_{12})\rightarrow \hat{\bi{d}}_1\cdot\hat{\bi{d}}_2.
	\label{eq:rzeroLimF}
	\end{equation}
	
	We can immediately see that for the collective effects to cause destructive interference and suppress the $\ket{-}$ rates we require $\mathrm{sign}\left[C\right]=\mathrm{sign}\left[\mathcal{F}\right]$. If this is not true, then one instead finds that the interference of the $\ket{-}$ rates is constructive. The magnitude of the interference term depends on: the size of the renormalized monomer detuning $\Delta$ compared to the renormalized dipole-dipole coupling $C^{\prime}$; the geometry of the dimer though the cross function $\mathcal{F}$ and coupling $C^{\prime}$; the bare monomer rates $\gamma_j(\omega)$ and the phonon coupling strengths through the factor $\kappa_1\kappa_2$ and renormalization of the monomer transition rates. 
	
	The monomer electronic transition rates, $\Gamma_j^{\mu}(\omega)$ in (\ref{eq:InitialMinusPhotonRate}) are discussed in \ref{App:DarkMonomer}. These cannot be derived analytically and the usual approximation would be to assume that the photon spectral density is flat for frequencies near to the eigenfrequencies of the system~\cite{nazir2016modelling,scerri2017method}. This is the flat spectral density approximation (FSDA). If we used this approximation we would find that $\Gamma_j^{\mathrm{A}}(\omega)\rightarrow\Gamma_j^{\mathrm{A},\mathrm{FSDA}}(\omega)=\gamma_j(\omega)N(\omega)$ and $\Gamma_j^{\mathrm{E}}(\omega)\rightarrow\Gamma_j^{\mathrm{E},\mathrm{FSDA}}(\omega)=\gamma_j(\omega)\left[1+N(\omega)\right]$. Hence, under the FSDA the phonon coupling does not affect the electronic transition rates. It is therefore not surprising that the rates derived using the FSDA are equivalent to those derived in the limit of weak phonon coupling. For strong phonon couplings this approximation quickly breaks down~\cite{maguire18environmental}, and we show this explicitly in Figure~\ref{figA1} in \ref{App:DarkMonomer}. Instead of making the FSDA, and because the effect of increasing the phonon coupling is to increase the lifetime of the monomers, we write
	\numparts
	\begin{equation}
	\Gamma_j^{\mu}(\omega)=\zeta^{\mu}_j(\omega)\Gamma_j^{\mu,\mathrm{FSDA}}(\omega),
	\label{eq:ElecRateE}
	\end{equation}
	\endnumparts
	where the $\zeta^{\mu}_j(\omega)$ have values $0\le \zeta_j^{\mu}(\omega)\le 1$, tending towards zero for increasing phonon coupling strength. Deviations from unity measure the error in assuming that the photon spectral density can be regarded as being flat, i.e. in assuming that there is no influence on optical emission rates by the vibrational environments. A similar result has been recently found in the context of mapping the vibrational environment onto a collective coordinate~\cite{maguire18environmental}.
	
	The functional form of $\zeta^{\mu}_j(\omega)$ is dependent on the choice of phonon spectral density, $J(\omega)$. An exact solution can only be derived for simple spectral densities. However, an approximate but analytic solution can be found for many spectral densities. We give this approximate expression (\ref{eq:zetafunctions}) in \ref{App:DarkMonomer}. In Figure~\ref{figA1} of \ref{App:DarkMonomer}, we plot the approximated and exact expressions for $\zeta^{\mu}_j(\omega)$ for a simple spectral density (where the exact solution exists) for a range of phonon couplings to illustrate that the approximate solution works well in this case. Throughout this paper, we use the spectral density
	\begin{equation}
	J(\omega)=A\omega^3\mathrm{exp}\left[-\left(\omega/\omega_c\right)^2\right],\label{eq:SD}
	\end{equation}
	for which there is no exact expression of $\zeta^{\mu}_j(\omega)$. Therefore, we assume the approximated expression (\ref{eq:zetafunctions}) is accurate for this spectral density.
	
	We can finally summarize the $\ket{-}$ absorption and emission rates by writing
	\begin{eqnarray}
	\gamma_-^{\mathrm{A}}=\Gamma_-^{\mathrm{A}}N(\delta_-),\\
	\gamma_-^{\mathrm{E}}=\Gamma_-^{\mathrm{E}}\left[1+N(\delta_-)\right],
	\end{eqnarray}
	where the rate coefficients are
	\begin{eqnarray}
	\Gamma_-^{\mu}=\sin^2\frac{\chi}{2}\zeta^{\mu}_1(\delta_-)\gamma_1(\delta_-)&+\cos^2\frac{\chi}{2}\zeta^{\mu}_2(\delta_-)\gamma_2(\delta_-)\nonumber\\
	&-\kappa_1\kappa_2\sin\chi\sqrt{\gamma_1(\delta_-)\gamma_2(\delta_-)}\mathcal{F},
	\label{eq:MinusPhotonRate}
	\end{eqnarray}
	for $\mu=\mathrm{A},\mathrm{E}$. 
	
	\subsection{The dark state}
	
	To make the $\ket{-}$ a dark state, we need to minimize the rate coefficient (\ref{eq:MinusPhotonRate}). To do so it is useful to define the dipole magnitude ratio, $z\equiv\left|\bi{d}_2/\bi{d}_1\right|$ which, under our definitions of monomer-1 and monomer-2, can take the values $0< z\le 1$. We identify two parameter regimes where the emission rate is minimized by two distinct mechanisms.
	
	The first mechanism occurs when the monomer detuning is much greater than the  dipole-dipole coupling ($\Delta\gg C^{\prime}$). From (\ref{eq:MinusES}) we see that this causes the antisymmetric eigenstate to localize onto monomer-2 ($\ket{-}\approx\ket{2}$) which is reflected in the emission and absorption rates of the $\ket{-}$ because
	\begin{equation}
	\lim_{\Delta\gg C^{\prime}}\Gamma_-^{\mu} \rightarrow \zeta^{\mu}_2(\delta_2^{\prime})\gamma_2(\delta_2^{\prime}),
	\label{eq:RateLocalize}
	\end{equation}
	and these are the respective rates for monomer-2. For the large detuning and small dipole magnitude ratio necessary to achieve localization, the emission rate of monomer-2 is small because $\gamma_2(\delta_2^{\prime})\propto z^2\left(\delta_1^{\prime}-\Delta\right)^3$ (see (\ref{eq:BETR})). Therefore, very heterogeneous dimers can lead to a dark $\ket{-}$. This mechanism does not make use of quantum effects, but simply localizes the eigenstate onto an already relatively optically inactive monomer. Therefore, these heterodimers cannot create perfect dark states without completely decoupling monomer-2 from the photon field. This is close to the classical picture of a dimer consisting of a donor (monomer-1) and acceptor (monomer-2) molecule.
	
	The second mechanism maximizes the destructive interference term of (\ref{eq:MinusPhotonRate}). This therefore requires $C^{\prime}\gg\Delta$ (so that $\sin\chi\approx 1$) which is achieved for homodimers because these have $(z,\Delta)\rightarrow (1,0)$. This can be more useful than the first mechanism because we can produce a highly (and in some ideal situations, perfectly) decoupled eigenstate but still have large dipole-dipole couplings. Since the phonon transition rates between the $\ket{+}$ and the $\ket{-}$ scale as $\sim C^{\prime 2}$, this also enables easier access to the dark state.  The drawback of this mechanism is that the interference is easily suppressed by strong phonon couplings and sub-optimal monomer orientations; this is in contrast to dark states in heterodimers, which are largely unaffected. This can can be inferred from
	the interference term in (\ref{eq:MinusPhotonRate}), which is proportional to $\kappa_1\kappa_2$ and $\mathcal{F}$, and so away from ideal geometries ($\mathcal{F}=1$) and very weak phonon couplings ($\kappa_j\approx 1$) the destructive interference is reduced.

	\subsection{Minimization of the emission rate}

\begin{figure}[ht!] 
		\centering
		\includegraphics[width=1\linewidth]{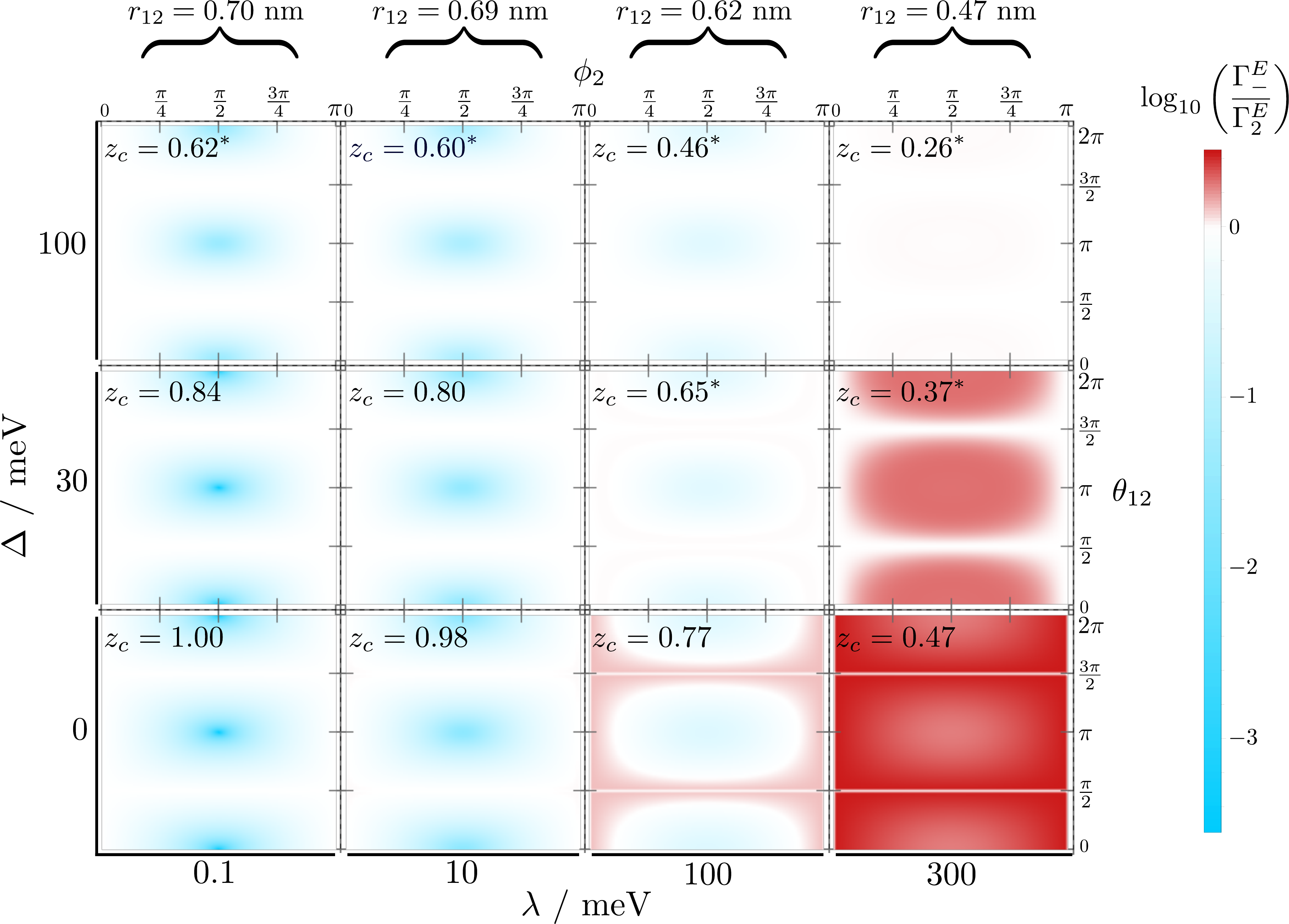} 
		\caption{Relative photon emission rate of the antisymmetric eigenstate $\ket{-}$ compared to the lower energy monomer in the dimer. The color axis is the logarithm of the ratio between the $\ket{-}$ and monomer-2 emission rates. Each subplot corresponds to a fixed reorganization energy and detuning; within each subplot we show the rate ratio as a function of angles $\phi_2$ and $\theta_{12}$, but keep $\phi_1=\frac{\pi}{2}$ fixed. Therefore, maximal destructive interference, $\mathcal{F}=1$, occurs for $\phi_2=\frac{\pi}{2}$ along with $\theta_{12}=0,\pi,2\pi$. In each subfigure, we keep the unnormalized lifetime of monomer-1 fixed at $\left(3\pi\right)/\left(d_1^2\delta_1^{\prime 3}\right)=5~\mathrm{ ns}$. This then fixes $d_1=0.15e$ nm in all plots, where $e$ is the electron charge. We should note that due to phonon renormalization, the measured lifetime of monomer-1 will be larger by a factor of $1/\zeta_1^{\mathrm{E}}(\delta_1^{\prime})$. Additionally, in each subfigure we fix the maximum value of the renormalized coupling to $C^{\prime}=\left(100\ \mathrm{meV}\right)z_c$ (in the ideal orientation) by varying the monomer separation, indicated above each column. Finally, we keep the polaron frame energy splitting of monomer-1 fixed at $\delta_1^{\prime}=2.8~\mathrm{ eV}$ and create the detuning by decreasing $\delta_2^{\prime}$. We assume that photon temperature has negligible effect. We choose identical phonon spectral densities for the two baths given by (\ref{eq:SD}) and keep $\omega_c=0.3~\mathrm{eV}$ fixed, which gives a realistic range of vibrational modes in organic molecules~\cite{sowa2018beyond}. Both phonon baths are at identical temperature, $T_{\mathrm{pn}}=300$~K. These phonon bath properties are used throughout the paper.} 
		\label{fig2}
	\end{figure}

	In Figure~\ref{fig2} we explore the dependence of the $\ket{-}$ emission rate on detuning, orientation and phonon coupling strength using the complete expression for the dipole-dipole coupling given by (\ref{eq:DipoleDipoleCoupling}). We plot the logarithm of the $\ket{-}$ emission rate, $\Gamma_-^{\mathrm{E}}$, divided by the emission rate of monomer-2, $\Gamma_2^{\mathrm{E}}=\zeta_2^{\mathrm{E}}(\delta_2^{\prime})\gamma_2(\delta_2^{\prime})$, where $\delta_2^{\prime}=\delta_1^{\prime}-\Delta$ and we fix $\delta_1^{\prime}=2.8$~eV. We are therefore using the lower energy monomer as a benchmark to test the decoupling. The coordinate system of the dipoles is defined in Figure~\ref{fig1}(b).
	
	If the emission rate of the $\ket{-}$ is larger than that of monomer-2 then this does not necessarily mean that the power output of the dimer will also be lower. This is explored in Section~\ref{Sec6}. In Figure~\ref{fig2}, we fix $\phi_1=\frac{\pi}{2}$ in all subfigures so that for all orientations $\mathrm{sign}\left[C\right]=\mathrm{sign}\left[\mathcal{F}\right]$ and therefore the interference is alwayy destructive. Additionally, we choose the value of $z$ such that the $\ket{-}$ emission rate (\ref{eq:MinusPhotonRate}) is minimized in the ideal orientation - i.e. in each subfigure we use the same value of $z$ for all orientations. The ideal orientation for maximizing destructive interference is dipoles parallel or anti-parallel to each other, and perpendicular to the interconnecting vector so that $\mathcal{F}=1$ (i.e. $(\phi_1,\phi_2,\theta_{12})=(\frac{\pi}{2},\frac{\pi}{2},n\pi)$ for $n=0,1,2$). Figure~\ref{fig2} then allows us to determine the effect of imperfect orientation.
	
	We keep the renormalized coupling fixed at $C^{\prime}=\left(100\ \mathrm{meV}\right)z$ and study a wide range of detunings. Therefore, we explore both the homodimer parameter regime where destructive interference is important ($C^{\prime}>\Delta$), as well as the heterodimer parameter regime where instead localization is important ($C^{\prime}\lesssim \Delta$). For parameters where the interference term is strongly suppressed the only minimum in the emission rate occurs at $z=0$, even in the optimal orientation. In Figure~\ref{fig2}, these situations are indicated with an asterisk ($\ast$) and we instead choose a $z$ value by using 
	\begin{equation}
	z_{\mathrm{c}}^{\ast}=\tan\frac{\chi}{2}\frac{\kappa_1\kappa_2}{\zeta^{\mathrm{E}}_2(\delta_-)}\mathcal{F},
	\end{equation} 
	derived in \ref{App:AnalyticMin}. This expression gives the value of $z$ for which the destructive interference minimizes the emission rate (\ref{eq:MinusPhotonRate}) in the simplified case where $C^{\prime}$ does not depend on dipole strength, an assumption made in~\cite{fruchtman2016photocell}. 
	
	Figure~\ref{fig2} clearly illustrates that the destructive interference is larger for detunings much smaller than the renormalized coupling, as evident in the bottom two rows. Additionally, we can see the extreme sensitivity of the emission rate to orientation and phonon coupling strength: As a guide, reorganization energies in organic molecules are typically of the order $100~\mathrm{meV}$. We also see in Figure~\ref{fig2} that to form dark states at strong phonon couplings it is more beneficial to have heterodimers ($\Delta>C^{\prime}$). As expected in these cases, the $\ket{-}$ emission is equal to the benchmark emission, which would be small due to the reduced $z$ and $\delta_2^{\prime}$ values. At strong coupling in homodimers ($\Delta=0$) we see that the emission rate of the $\ket{-}$ can exceed that of monomer-2. This is because with suppressed destructive interference, the partial localization of the eigenstate over the more optically active monomer-1 dominates. In fact, at very strong couplings and small detunings it is preferable to orientate the dipoles such that the coupling $C^{\prime}$ is reduced much smaller than $\Delta$ to localize onto the lower energy monomer, so forming a heterodimer. This is not possible in the bottom row where the detuning is zero, except for $\theta_{12}=\frac{\pi}{2},\frac{3\pi}{2}$ for which $C^{\prime}=0$.

	\section{Spectra}\label{Sec5}
\begin{figure}[ht!] 
		\centering
		\includegraphics[width=1\linewidth]{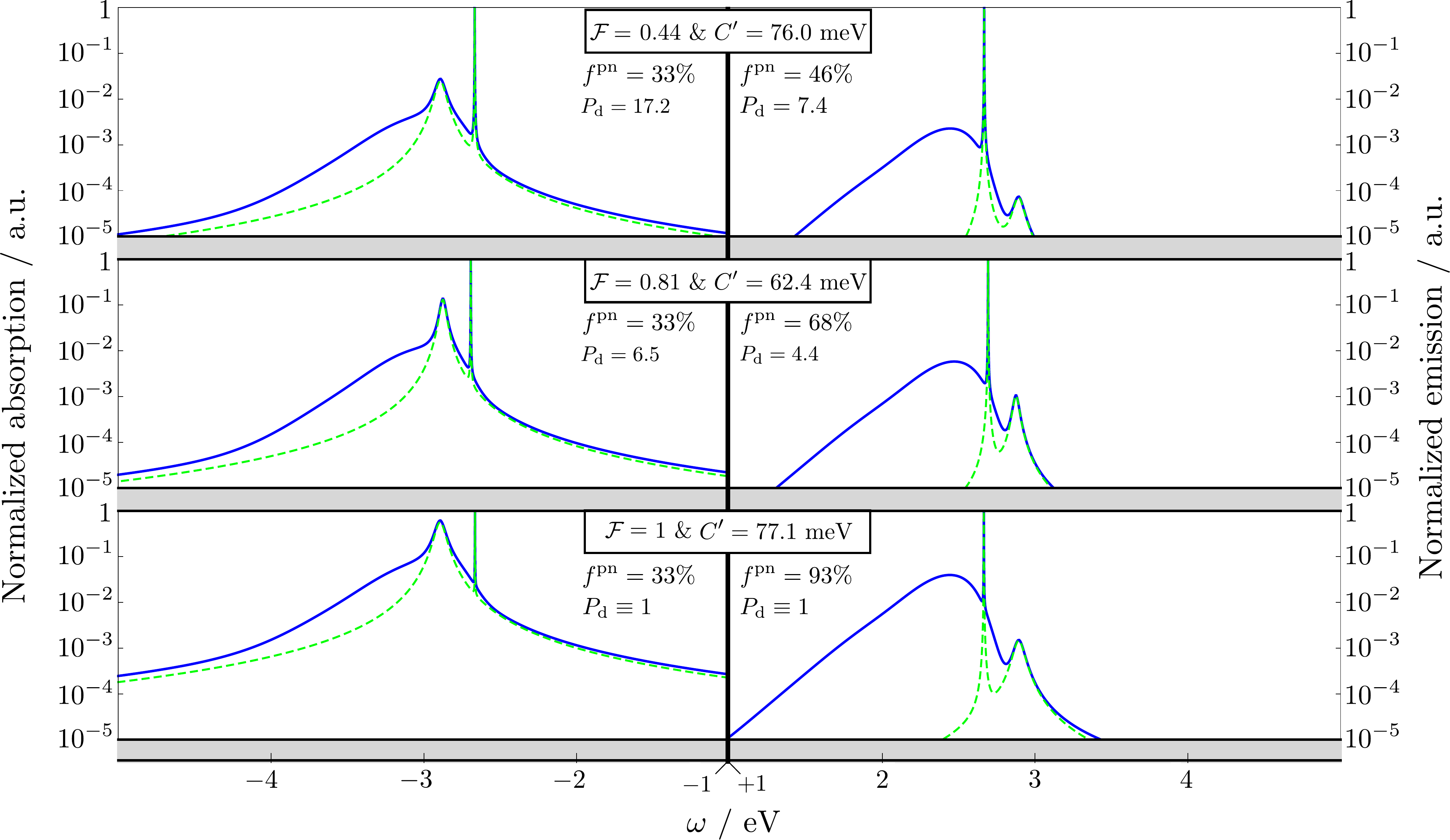} 
		\caption{Absorption and emission spectra for the homodimer in different orientations. We plot the spectra with (solid blue) and without (dashed green) the phonon sideband. We explain in \ref{App:Spectra} how to remove the phonon sideband from the calculated spectra; we note that phonons do play a role in all spectra. Monomer-1 and monomer-2 have renormalized energies $\delta_1^{\prime}=\delta_2^{\prime}=2.8$~eV with $\lambda=100$~meV for both environments. All other parameters are identical to those used in Figure~\ref{fig2}, unless explicitly stated otherwise. The orientations in each row from bottom to top, written in the form $(\phi_1,\phi_2,\theta_{12})$, are $(\frac{\pi}{2},\frac{\pi}{2},0)$, $(\frac{\pi}{2},\frac{\pi}{2},0.2\pi)$ and $(0.6\pi,0.3\pi,0.2\pi)$. The corresponding cross functions and dipole-dipole couplings for $z=0.735$ are given in the figure. For each spectra we highlight certain parameters: $f^{\mathrm{pn}}$ is the fraction of absorption or emission in the phonon sideband and $P_{\mathrm{d}}$ is the relative intensity of emission and absorption from the dark state, $\ket{-}$. $P_{\mathrm{d}}$ is measured relative to the ideal dimer, whose spectra are in the bottom row, and calculated using the area of the peaks (positioned at the environment-renormalized frequency of the dark state, see \ref{App:PhotonLiouvillian} and \ref{App:PhononLiouvillian}) above the full-width-half-maximum line.} 
		\label{fig3}
	\end{figure}

	Producing homodimers that maximize destructive interference requires precise control of the monomer orientations. This is especially true at strong phonon couplings where the interference has already been suppressed. To highlight experimentally detectable signatures of dark state formation in a homodimer, we plot theoretically calculated absorption and emission spectra in Figure~\ref{fig3}. In \ref{App:AsymDimer} we plot spectra for the heterodimer. 
	
	We choose a realistic molecular phonon coupling strength for both environments, $\lambda_j=100$~meV for $j=1,2$ and keep the phonon renormalized dipole-dipole coupling at $C^{\prime}/z=100$~meV in the optimum orientation. In each row of Figure~\ref{fig3} we choose different orientations of the dipoles. The lowest row corresponds to the orientation which optimizes dark state formation (signified by $\mathcal{F}= 1$) and progressively higher rows correspond to less desirable geometries (signified by $\mathcal{F}< 1$). The specific angles are detailed in the caption. We use the same dipole magnitude ratio, $z=0.735$, in each row. This value minimizes the $\ket{-}$ emission rate, i.e. creates the dark state in the homodimer, for the optimum orientation. In this way, the lowest row corresponds to the dimer one should aim to produce in order to maximize decoupling from the photon field, and the higher two rows show realistic experimental deviations from this perfect scenario. We note that the minimizing value of $z$ is not 1 (i.e. the dimer is not a completely homogeneous) and this is because of the strong phonon coupling.
	
	In each subfigure of Figure~\ref{fig3} we plot the spectra with and without the phonon sideband (as solid blue and dashed green lines respectively). Using these plots we are also able to calculate the fraction of emission or absorption which occurs through the phonon sideband numerically, which is the parameter $f^{\mathrm{pn}}$: The specifics of this calculation, and how we find the spectra, are discussed in detail in \ref{App:Spectra}. 
	
	From Figure~\ref{fig3} we can identify two different signatures that a dark state has formed in the homodimer. Firstly, the darker the $\ket{-}$, the more reliant the dimer is on phonons for emission. This is reflected by the trend in $f^{\mathrm{pn}}$. For these strong phonon and dipole-dipole couplings, excitations are mostly transferred to the $\ket{-}$ before they are immediately re-radiated from the $\ket{+}$. Therefore, emission can only occur after phonon transfer out of the dark state, which becomes increasingly more true the darker the $\ket{-}$ is. Crucially, the excitations are trapped without phonon transfer. This trend is not seen in the absorption spectra because the dimer can still be excited through the $\ket{+}$ even if the $\ket{-}$ is dark. In this case, there is no absolute blocking of absorption in the absence of phonon processes. As we show in \ref{App:fpn}, one can derive expressions for $f^{\mathrm{pn}}$ which are exact for absorption and approximate for emission. We also demonstrate the accuracy of these expressions in Figure \ref{figD2}. For the homodimer, one finds that (for $\kappa_1=\kappa_2\equiv\kappa$) the fractional absorption and emission in the phonon sideband are
	\begin{eqnarray}
	f^{\mathrm{pn}}_{\mathrm{A}}= 1-\kappa^2,\label{eq:fpnA}\\
	f^{\mathrm{pn}}_{\mathrm{E}}\approx \frac{1-\kappa^2}{1-\kappa^2\mathcal{F}\mathrm{sign}\left[C\right]}.\label{eq:fpnESymmetric}
	\end{eqnarray}
	For the parameters used in Figure \ref{fig3} the value of $1-\kappa^2=0.33$, in agreement with the numerical values calculated from the spectra. Indeed, (\ref{eq:fpnESymmetric}) allows one to determine $\mathcal{F}$ experimentally for a homodimer by measuring $f^{\mathrm{pn}}_{\mathrm{E}}$.
	
	The second parameter which helps to identify dark state formation in a homodimer is the total amount of light that is absorbed and emitted by the dark state, $P_{\mathrm{d}}$. As expected, the total absorption and emission from the dark state dramatically increases as the destructive interference weakens due to poor geometry.

	\section{Power output}\label{Sec6}

\begin{figure}[ht!] 
		\centering
		\includegraphics[width=1\linewidth]{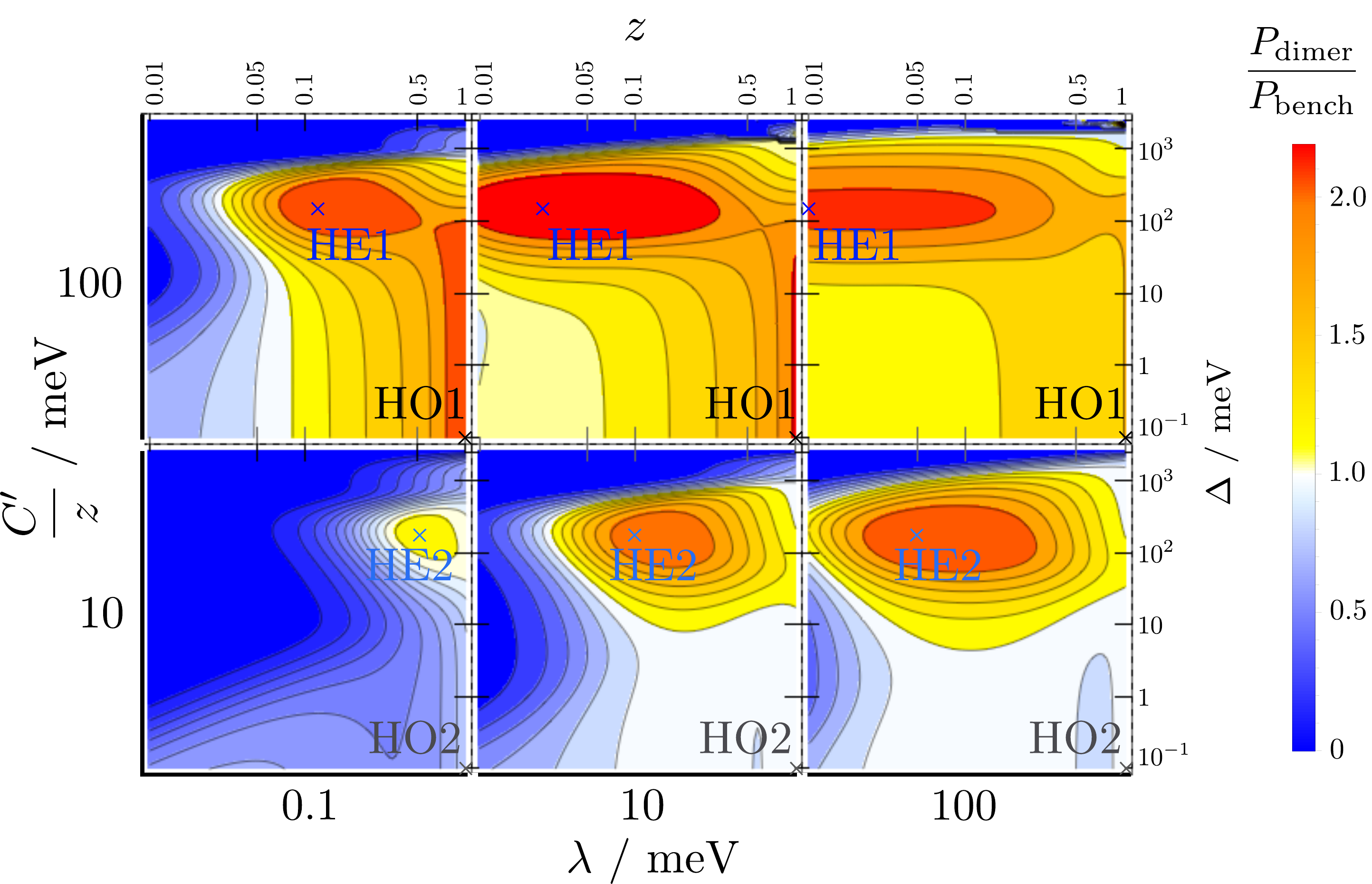} 
		\caption{Power output of the dimer across a range of system parameters with extraction at $\gamma_x=100$~neV. As before, we keep the two phonon baths identical and fix the (unnormalized) lifetime of monomer-1 to be 5 ns with a renormalized splitting of 2.8~eV. In each subfigure we vary the energy splitting of monomer-2 from $2.8$~eV to $0$ and $z$ from 0 to 1. In order to keep both the dipole-dipole coupling and monomer-1 lifetime fixed at different reorganization energies, the monomer separation is changed in each contour plot. The separations in nm (reading first from left to right across the top row and then repeating with the bottom) are 0.69, 0.61, 0.47; 1.49, 1.32, 1.02. Following~\cite{higgins2017quantum}, in each calculation the transfer rate within the trap, $\gamma_t$ is optimized to maximize power output. The photon environment is at the temperature of the solar surface, $T_{\gamma}=6000$~K.} 
		\label{fig4}
	\end{figure}

	Finally, we have calculated the power output of the dimer, shown in Figure~\ref{fig4}; we use six panels for different values of dipole coupling $C^\prime$ and phonon coupling $\lambda$, and within each panel vary $z$ and $\Delta$. To calculate the power output we formally introduce the idealized load model alluded to in Figure~\ref{fig1}(c). Following~\cite{higgins2017quantum,scully2010quantum}, we measure the power output of the dimer by coupling it to a distinct two level system called a trap, which serves as the idealized load. The extraction process is modelled by adding an incoherent process which moves excitations from a chosen eigenstate in the dimer (which we will choose to be the $\ket{-}$) to the ground state, whilst simultaneously exciting the trap, all at a rate $\gamma_{\mathrm{x}}$. There is also an incoherent process which causes transfer from the excited trap-state $\ket{\alpha}$ to the ground trap-state $\ket{\beta}$ at rate $\gamma_{\mathrm{t}}$, and this is always optimized to give maximal power output. The voltage and current through the trap at inverse temperature $\beta_{\mathrm{t}}=\left(k_BT_{\mathrm{t}}\right)^{-1}$ are then defined as
	\begin{eqnarray}
	V=\delta_{\mathrm{t}}-\frac{1}{\beta_{\mathrm{t}}}\mathrm{ln}\frac{P_{\alpha}}{P_{\beta}},\label{eq:Voltage}\\
	I=e\gamma_{\mathrm{t}} P_{\alpha},
	\end{eqnarray}
	where $\delta_{\mathrm{t}}$ is the trap energy splitting, $e$ is the electron charge and the $P_{i}$ for $i=\alpha,\beta$ are the trap populations in the steady state. The trap Hamiltonian is $H_{\mathrm{t}}=\delta_{\alpha}\ket{\alpha}\bra{\alpha}+\delta_{\beta}\ket{\beta}\bra{\beta}$, where $\delta_{\mathrm{t}}=\delta_{\alpha}-\delta_{\beta}$ is always set equal to the energy splitting of the state from which the extraction is occurring. The extraction and transfer processes are added phenomenologically with Lindblad operators~\cite{higgins2017quantum}. For a full description see \ref{App:RC}. Other models for the extraction process and power generation are also possible~\cite{gelbwaser2017thermodynamic}.
	
	Instead of plotting absolute power (which is shown in \ref{App:AbsolutePower}), Figure~\ref{fig4} shows the power output of the dimer compared to a benchmark. The benchmark is the total power that would be extracted from the monomers in the dimer if they were completely independent and coupled to separate traps. In the benchmark, the extraction rate from either monomer is equal to the rate that we extract from the dark state in the dimer. Therefore, if the dimer produces more power than the benchmark, then we can say with certainty that using the dimer conveys a definite advantage because overall the dimer has only half the extraction rate of the benchmark. In Figure~\ref{fig4}, the power output is plotted as a function of dipole magnitude ratio $z$ and detuning $\Delta$, over dipole-dipole couplings and phonon couplings ranging from the weak to strong regimes. The extraction rate is fixed at $\gamma_x=100$~neV. Indicated on the plots are four power maxima which correspond to forming dark states in the homogeneous and heterogeneous dimers.
	
	The four power maxima are labelled `homodimer maximum-X' (HOX) and  `heterodimer maximum-X' (HEX) for $\mathrm{X}=1,2$ which label the strong and weak dipole-dipole coupling regimes respectively. The label of the maximum refers to the detuning and relative dipole strengths of the monomers in the dimer and therefore, as discussed in Section~\ref{Sec4}, the mechanism by which the $\ket{-}$ is optically darker. We see that the HO1 has a higher power output than the HO2 because of stronger dipole-dipole coupling which increases the destructive interference and that, as expected, strong phonon couplings quench this. Since the monomer detunings are zero at the HOX, the eigenstate detuning $\eta=\sqrt{\Delta^2+C^{\prime 2}}= C^{\prime}$. To maximize overall phonon transfer from the $\ket{+}$ to the $\ket{-}$ the eigenstate detuning must optimized. The detuning must be sufficiently large to suppress excitation via phonon transfer back out of the dark state, but smaller than the cut-off frequency of the vibrational baths $\omega_c$ so that the modes that are most strongly coupled are used to complete the transfer. As before, we use a cut-off frequency found in realistic molecules $\omega_c=0.3$ eV \cite{sowa2018beyond} therefore at the HOX, $\eta< \omega_c$. Additionally, in the polaron frame the phonon rates scale with $C^{\prime 2}$ (see \ref{App:PhononLiouvillian}). Therefore, in addition to being darker, phonon transfer is also much larger at the HO1 than the HO2. 
	
	As discussed in Section \ref{Sec4}, at the HEX the $\ket{-}$ is localized onto monomer-2, which requires small $z$ and $\delta_2^{\prime}$. With these parameters monomer-2 is also relatively optically inactive. This localization also implies that $C^{\prime}\ll \Delta$ and therefore the eigenstate splitting is dominated by the monomer detuning, i.e. $~{\eta=\sqrt{\Delta^2+C^{\prime 2}}\approx\Delta}$. The sacrifice made with the small dipole strength means that phonon transfers, which scale with $C^{\prime 2}$ are suppressed. Therefore, the power maxima occur at the eigenstate detuning that will maximize phonon transfer from the $\ket{+}$ to the $\ket{-}$. For our chosen cut-off frequency, phonon transfers maximize for eigenstate detunings $\approx0.2$ eV which explains the location of the HEX. We show important rates in understanding the power outputs in \ref{App:Rates}.
	
\begin{figure}[ht!] 
		\centering
		\includegraphics[width=1\linewidth]{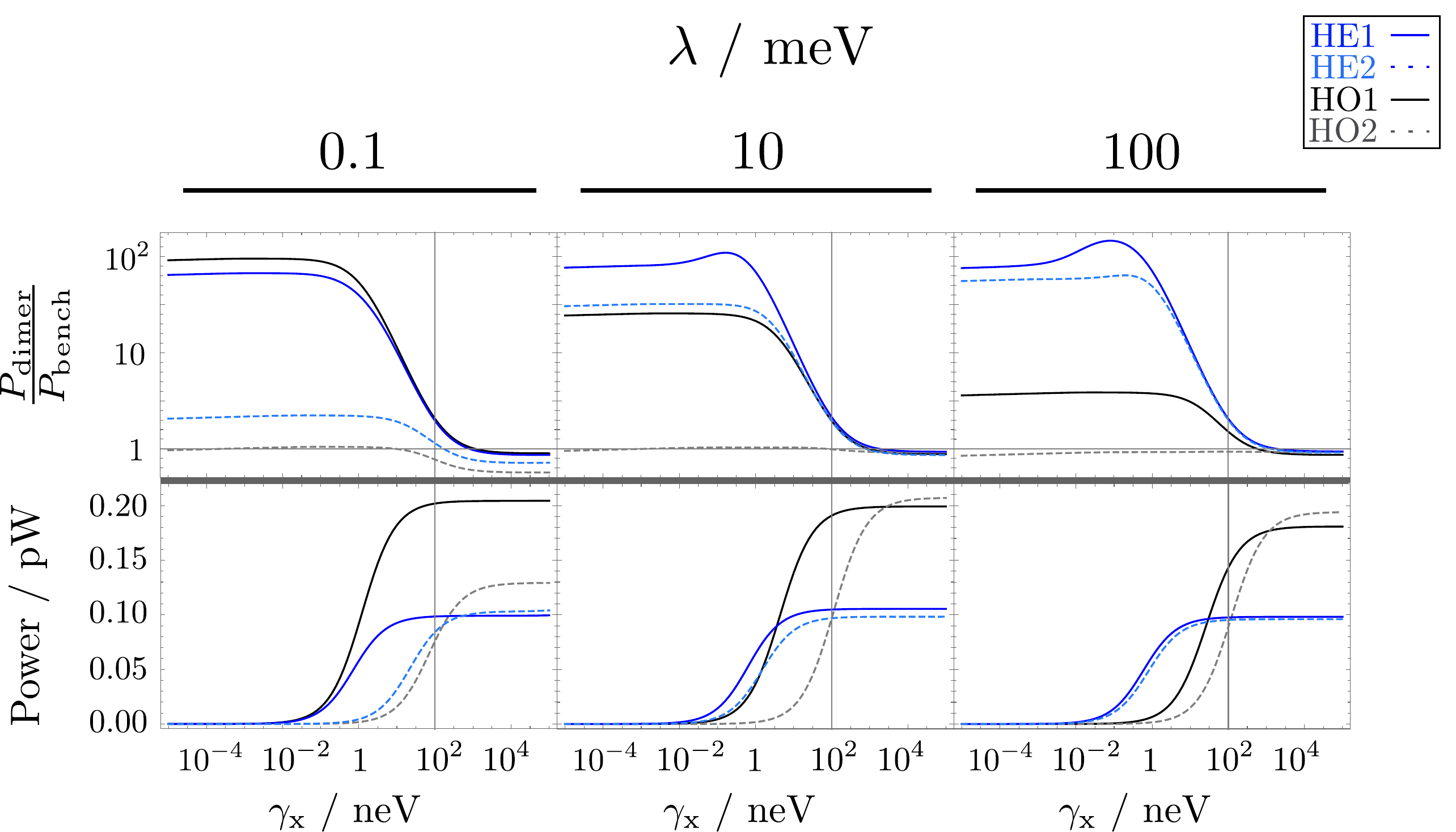} 
		\caption{Absolute and benchmarked power output of the dimer at each of the HOX and HEX as a function of extraction rate. The lines in each plot have the detunings and dipole magnitude ratios which produce the maxima identified in Figure~\ref{fig4}. The vertical line labels the extraction rate used in Figure~\ref{fig4}.}  
		\label{fig5}
	\end{figure}

	In Figure~\ref{fig5} we plot both the benchmarked power and the absolute power at each of these maxima as a function of extraction rate. Although absolute power output is always greater for faster extraction to the trap, it does not necessarily mean that a real device would operate in these high extraction regimes. This is discussed in the supplementary material of~\cite{higgins2017quantum}. There it is argued that if one assumes a simple linear relationship between the total extraction rate and the average number of traps adjacent to any light harvesting dimer. Natural photosynthetic systems commonly contain a few hundred antennae for each reaction centre~\cite{croce2014natural}, which leads to a relatively small transfer rate. It can therefore be important to improve power output at low extraction rate, which is the regime in which dark state protection typically works well. 
	
	From Figure~\ref{fig4} and Figure~\ref{fig5} we identify that the strengths of the dipole-dipole coupling and phonon couplings should dictate whether the target dimer should be homogeneous or heterogeneous. The power output achieved with homodimers can be large, however, it requires either extremely large dipole-dipole couplings ($C^{\prime}>100$~meV) or extremely weak phonon couplings $\lambda \approx 0.1$~meV. Such strong dipole-dipole couplings are only achievable with unrealistically short monomer lifetimes and/or separations. Organic monomers typically have reorganization energies close to $100$~meV and in this regime aiming for the HOX is not ideal. On the other hand, the HE1 produces a high power output which is relatively unaffected by increasing phonon coupling. This reflects the fact that the HE1 is not reliant on an ever-increasingly quenched destructive interference for low $\ket{-}$ emission. In the HE2, increasing phonon coupling actually increases power output because, at this small dipole-dipole coupling, the power bottleneck is the phonon transfer rate from the bright to the dark state.
	
	In Figure~\ref{fig5} we see that for large extraction rate the advantage conferred by dark state protection is lost, for all of the HEX and HOX. This is because extraction must, beyond some threshold rate, always occur much more quickly then any exciton recombination. There is, however, also a low extraction rate regime where the power output of the benchmark slightly exceeds that of the HO2 dimer. This happens for strong phonon coupling ($\lambda=100$~meV) which destroys the destructive interference effect for this dark state. Since there is then no dark state advantage, and the benchmark has overall twice the extraction, the benchmark has a higher power output than the dimer.

	\section{Conclusion}\label{Sec7}
	Dimers are the relatively simple molecules that can form the building blocks of larger more sophisticated light harvesting complexes that could form the components of organic solar cells~\cite{creatore2013efficient}. Therefore, a detailed understanding of the energy transfer processes in these systems subject to realistic environmental constraints is important. Here we have developed a theory for energy transfer in a dimer which significantly improves upon previous models. More importantly, we have tailored the model to better suit biologically inspired dimers where the coupling to vibrational modes can be strong and a polaron theory is needed to capture the details of the dynamics. 
	
	We have found that there can be an enhanced power output for the homodimers which exploit the quantum effect of delocalization, resulting in destructive interference of the optical emission rate. However, this is extremely sensitive to orientation, phonon coupling and the ratio of detuning with dipole-dipole coupling. For strong phonon coupling, which is close to experimental reality, it can be beneficial to instead design a heterodimer that localizes the eigenstates onto the monomers. The optimal dimer choice also depends on how quickly excitons are extracted into the electrical circuit and there are additional, distinct challenges in constructing the best homogeneous or heterogeneous dimer. 
	
	Engineering an heterodimer (HEX) requires careful pairing of monomers to localize the $\ket{-}$ onto a relatively dark monomer-2, whilst ensuring that there is sufficient phonon transfer into the $\ket{-}$ so that it is populated. However, this is mitigated by there being a vast number of monomers to choose from, as shown in the supplementary material of \cite{fruchtman2016photocell}. On the other hand, creating a homodimer (HOX) simply requires the use of identical monomers with weak vibrational environments. However, the dipoles of the monomers must be orientated with relatively high precision to maximize destructive interference, and we have identified experimentally detectable signatures of when this has been achieved. Since destructive interference is of little importance to heterodimers, their orientation can instead be utilized to fine-tune the magnitude of the dipole-dipole coupling. This enables one to find the optimal balance between localization and phonon transfer. In previous literature, there has been focus on producing dark states by maximizing the destructive interference using homodimers. Here we show that in organic solar cells, where the vibrational coupling is likely to be strong, this interference is significantly quenched, and it is instead better to produce heterodimers where the eigenstate localizes onto a relatively optically dark monomer.

	\section*{Acknowledgements}
	We thank Jonathan Keeling, Alexandre Coates, William Brown, Aidan Strathearn and Dale Scerri for useful discussions. D.R. acknowledges studentship funding from EPSRC under grant no.~EP/L015110/1. E.M.G. thanks the Royal Society of Edinburgh and the Scottish Government for support.

	\clearpage
	\appendix
	\section*{Appendices}
	\section{Emission and absorption rates from the antisymmetric eigenstate}\label{App:DarkEmission}
	To derive the absorption and emission rates of the $\ket{-}$ we start from the ECF definition (\ref{eq:ECF}). We first partition the photon interaction Hamiltonian (\ref{eq:PhotonPolaronInt}) into system and environment operators as $\tilde{H}_{\mathrm{IP}}^{\gamma}=\sum_{\alpha}A^{\gamma}_{\alpha} \otimes C_{\alpha}^{\gamma}$ where $A_{\mathrm{a}}^{\gamma}=\ket{0}\bra{+}$, $A_{\mathrm{b}}^{\gamma}=\ket{0}\bra{-}$, $A_{\mathrm{c}}^{\gamma}=A_{\mathrm{a}}^{\gamma\dagger}$ and $A_{\mathrm{d}}^{\gamma}=A_{\mathrm{b}}^{\gamma\dagger}$. The corresponding environment operators are $C_{\mathrm{a}}^{\gamma}=\rmi\sum_{\bi{q}\lambda}D^{(+)\dagger}_{\bi{q}\lambda}a^{\dagger}_{\bi{q}\lambda}$, $C_{\mathrm{b}}^{\gamma}=\rmi\sum_{\bi{q}\lambda}D^{(-)\dagger}_{\bi{q}\lambda}a^{\dagger}_{\bi{q}\lambda}$, $C_{\mathrm{c}}^{\gamma}=C_{\mathrm{a}}^{\gamma\dagger}$ and $C_{\mathrm{d}}^{\gamma}=C_{\mathrm{b}}^{\gamma\dagger}$ where $D^{(\pm)}_{\bi{q}\lambda}$ are given by (\ref{eq:EigenCoupling1}) and (\ref{eq:EigenCoupling2}). Then, from the expression for the general dissipator (\ref{eq:Dissipator}), one finds that the emission (E) and absorption (A) rates from the $\ket{-}$ to and from the ground state are given by
	\numparts
	\begin{eqnarray}
	\gamma_-^{\mathrm{E}}=2\Re\left[\Xi^{\gamma}_{\mathrm{bb}}(\delta_-)\right],\label{eq:DefinitionERates}\\
	\gamma_-^{\mathrm{A}}=2\Re\left[\Xi^{\gamma}_{\mathrm{dd}}(-\delta_-)\right].\label{eq:DefinitionARates}
	\end{eqnarray}
	\endnumparts
	In the following subsections we aim to calculate $\Xi_{\mathrm{bb}}^{\gamma}(\delta_-)$ and $\Xi_{\mathrm{dd}}^{\gamma}(-\delta_-)$. This first requires us to derive expressions for the two-time expectation functions over the environment space operators, which we do in the next subsection. Subsequently we must integrate these expressions over time.

	\subsection{Photon and phonon expectation values.}
	We first note that the environment space operators $C_{\alpha}^{\gamma}$ live in both the photon and phonon spaces, and that the expectation value in the ECF definition (\ref{eq:ECF}) is over both environments. The Born approximation allows us to separate the expectation value over both environments into a product of averages over single environments. Introducing the concise notations, 
	\begin{eqnarray}
	G_{i j}\equiv\left[\bi{d}_{i}\cdot \bi{u}_{\bi{q}\lambda}(\bi{r}_{i})\right]^*\left[\bi{d}_{ j}\cdot \bi{u}_{\bi{q}^{\prime}\lambda^{\prime}}(\bi{r}_{ j}) \right],\\
	B_{i j}(t)\equiv\langle B^{+}_{i}(t)B^{-}_{ j}(0) \rangle_{\mathrm{E},{\mathrm{pn}}}, 
	\end{eqnarray}
	where $i,j$ labels the monomers, we can write 
	\begin{eqnarray}
	\fl\langle C_{\mathrm{b}}^{\dagger}(t)C_{\mathrm{b}}(0)\rangle_{\mathrm{E}}=\sum_{\bi{q}\lambda}\sum_{\bi{q}^{\prime}\lambda^{\prime}}\Bigg[\sin^2\frac{\chi}{2}G_{11}B_{11}(t) + \cos^2\frac{\chi}{2}G_{22}B_{22}(t)-\nonumber\\
	\frac{1}{2}\sin\chi\left(G_{12}B_{12}(t)+G_{21}B_{21}(t) \right) \Bigg]\langle a^{\dagger}_{\bi{q}\lambda}(t)a_{\bi{q}^{\prime}\lambda^{\prime}}(0) \rangle_{\mathrm{E},\gamma},
	\label{eq:Photon22}
	\end{eqnarray}
	where in the interaction picture $a_{\bi{q}\lambda}(t)=a_{\bi{q}\lambda}\rme^{-\rmi\nu_{\bi{q}}t}$ and
	\begin{equation}
	B^{\pm}_{ j}(t)=\mathrm{exp}[\pm\sum_{\bi{k}}(g_{\bi{k}, j}b^{\dagger}_{\bi{k}, j}\rme^{\rmi\omega_{\bi{k}, j}t}-g^*_{\bi{k},j}b_{\bi{k}, j}\rme^{-\rmi\omega_{\bi{k}, j}t})/\omega_{\bi{k}, j}].
	\end{equation}
	As is shown in \cite{nazir2016modelling},  $\langle B^{+}_{i}(t)B^{-}_{ j}(0) \rangle_{\mathrm{E},{\mathrm{pn}}}=\langle B^{-}_{i}(t)B^{+}_{ j}(0) \rangle_{\mathrm{E},{\mathrm{pn}}}$, and therefore we can write that
	\begin{eqnarray}
	\fl\langle C_{\mathrm{d}}^{\dagger}(t)C_{\mathrm{d}}(0)\rangle_{\mathrm{E}}=\sum_{\bi{q}\lambda}\sum_{\bi{q}^{\prime}\lambda^{\prime}}\Bigg[\sin^2\frac{\chi}{2}G^*_{11}B_{11}(t) + \cos^2\frac{\chi}{2}G^*_{22}B_{22}(t)-\nonumber\\
	\frac{1}{2}\sin\chi\left(G^*_{12}B_{12}(t)+G^*_{21}B_{21}(t) \right) \Bigg]\langle a_{\bi{q}\lambda}(t)a^{\dagger}_{\bi{q}^{\prime}\lambda^{\prime}}(0) \rangle_{\mathrm{E},\gamma}.
	\label{eq:Photon44}
	\end{eqnarray}
	Aside from the averages over the photon operators, the expressions for the emission (\ref{eq:Photon22}) and absorption (\ref{eq:Photon44}) expectation values are largely identical. The thermal averages of the photonic operators are 
	\numparts
	\begin{eqnarray}
	\langle a^{\dagger}_{\bi{q}\lambda}a_{\bi{q}^{\prime}\lambda^{\prime}} \rangle_{\mathrm{E},{\gamma}}=\delta_{\bi{q}\bi{q}^{\prime}}\delta_{\lambda\lambda^{\prime}}N(\nu_{\mathrm{q}}),\\\langle a_{\bi{q}\lambda}a^{\dagger}_{\bi{q}^{\prime}\lambda^{\prime}} \rangle_{\mathrm{E},{\gamma}}=\delta_{\bi{q}\bi{q}^{\prime}}\delta_{\lambda\lambda^{\prime}}\left[1+N(\nu_{\mathrm{q}})\right].
	\end{eqnarray}
	\endnumparts
	
	We next want to evaluate $\sum_{\bi{q}\lambda}G_{i j}$ for $i, j=1,2$. This is completed by first taking the continuum limit, 
	\begin{equation}\sum_{\bi{q}}\rightarrow\frac{V}{(2\pi c)^3}\int_0^{\infty}\rmd \nu_{\mathrm{q}}\nu_{\mathrm{q}}^2\int_{\Omega_{\mathrm{q}}}\rmd \Omega_{\mathrm{q}},
	\end{equation} where $\rmd \Omega_{\mathrm{q}}$ is an infinitesimal solid angle. Then, by defining a coordinate system one can show that in this limit
	\begin{eqnarray}
	\fl\sum_{\bi{q}\lambda}G_{i j}=\sum_{\bi{q}\lambda}G^*_{i j}&\rightarrow\frac{1}{2\pi}\int_0^{\infty}\rmd \nu_{\mathrm{q}}
	\times \cases{\gamma_{ j}(\nu_{\mathrm{q}})&  for  $i= j$\\
		\sqrt{\gamma_{i}(\nu_{\mathrm{q}})\gamma_{j}(\nu_{\mathrm{q}})}\mathcal{F}(\nu_{\mathrm{q}}r_{ij})& for $i\neq j$\\}\label{eq:ContinuumLimitGij}\\
	\fl&\equiv \int_0^{\infty}\rmd \nu_{\mathrm{q}}F_{i j}(\nu_{\mathrm{q}}).\nonumber
	\end{eqnarray}
	where $\gamma_j(\omega)$ and $\mathcal{F}(x)$ are defined in (\ref{eq:BETR}) and (\ref{eq:FFunction}) respectively. This part of a similar derivation is discussed in detail in \cite{ficek2005quantum}.
	
	We also need the expectation values of the phonon displacement operators. Using standard results derived in \cite{nazir2016modelling} we have that:
	\begin{equation}
	B_{ij}(t)=\cases{\kappa_{ j}^2\rme^{\phi_{ j}(t)}& for $i= j$\\
		\kappa_{i}\kappa_{ j}& for $i\neq  j$\\},
	\label{eq:PhononTrace}
	\end{equation}
	where $\phi_j(t)$ is defined by (\ref{eq:PhononProp}).
	
	We now have the ingredients to calculate the ECFs. These require us to evaluate double integrals over photon frequency and time, which contain functions originating from both photon and phonon environments. Using the expressions for the expectation values, we find that these integrals are of the form
	\begin{equation}
	I_{i j}^{\pm}(\omega)=\int_0^{\infty}\rmd \nu_{\mathrm{q}}F^{\pm}_{i j}(\nu_{\mathrm{q}})\int_0^{\infty}\rmd t \rme^{i\left(\omega\pm\nu_{\mathrm{q}}\right)}B_{i j}(t),
	\label{eq:IpmdoubleInt}
	\end{equation}
	where we have defined $F_{i j}^+(\nu_{\mathrm{q}})=F_{i j}(\nu_{\mathrm{q}})N(\nu_{\mathrm{q}})$ and $F_{i j}^-(\nu_{\mathrm{q}})=F_{i j}(\nu_{\mathrm{q}})[1+N(\nu_{\mathrm{q}})]$. This enables us to succinctly  write the emission (\ref{eq:DefinitionERates}) and absorption (\ref{eq:DefinitionARates}) rates of the $\ket{-}$ as
	\begin{equation}
	\gamma_-^{{\mathrm{A}}/{\mathrm{E}}}=2\Re\left[\sin^2\frac{\chi}{2}I^{\pm}_{11}(\mp\delta_-)+\cos^2\frac{\chi}{2}I^{\pm}_{22}(\mp\delta_-)-\sin\chi I^{\pm}_{12}(\mp\delta_-)\right],
	\end{equation} 
	where we have used the fact that $I^{\pm}_{12}(\omega)=I^{\pm}_{21}(\omega)$.
	
	The integrals $I^{\pm}_{12}(\mp\omega)$ lead to the collective effects induced by the monomer coupling on the $\ket{-}$ rates with energy $\omega$, whereas  $I^{\pm}_{ j j}(\mp\omega)$ lead to the independent monomer rates for a monomer-$j$ with splitting $\omega$.

	\subsection{Monomer collective effects.}\label{App:DarkCollective}
	$I^{\pm}_{12}(\mp\omega)$ is calculated using the standard identity
	\begin{equation}
	\int_0^{\infty}\rmd y \rme^{\rmi\epsilon y}=\pi\delta(\epsilon)+\rmi\frac{\mathcal{P}}{\epsilon},
	\label{eq:CauchyIdentity}
	\end{equation}
	where $\delta(\epsilon)$ is the Dirac delta function and $\mathcal{P}$ indicates that the principal value of the subsequent integral should be taken. Using this it is straightforward to show that
	\begin{equation}
	I_{12}^{\pm}(\omega)= \pi\kappa_{1}\kappa_{2}\left[ F_{12}^{\pm}(\mp\omega)+\rmi\mathcal{P}\int_0^{\infty}d\nu_{\mathrm{q}}\frac{F_{12}^{\pm}(\nu_{\mathrm{q}})}{\omega\pm\nu_{\mathrm{q}}}\right].
	\label{eq:IJpm12}
	\end{equation}
	The real part of this quantity characterizes the changes to the eigenstate photon absorption and emission rates, and  the imaginary part describes the renormalization of the eigenfrequency $\delta_-$, which does not diverge and is given explicitly in \ref{App:PhotonLiouvillian}. Since these both originate from the dipole-dipole coupling between the monomers, they have a significant dependence on the orientations of the dipoles. The entire expression is renormalized by the phonon coupling strength of both monomers to their respective environments through $\kappa_1\kappa_2$.
	
	\subsection{Monomer phonon-renormalized electronic rates.}\label{App:DarkMonomer}Finally, we need to calculate the independent rates for each monomer. The relevant integrals are 
	\begin{equation}
	I_{ j j}^{\pm}(\omega)=\kappa_{ j}^2\int_0^{\infty}\rmd \nu_{\mathrm{q}}F_{ j j}^{\pm}(\nu_{\mathrm{q}})\int_0^{\infty}\rmd t \rme^{\rmi\left(\omega\pm\nu_{\mathrm{q}}\right)t}\rme^{\phi_{ j}(t)}.
	\label{eq:startingpoint}
	\end{equation}
	Unlike in the collective case, we see from (\ref{eq:PhononTrace}) that the renormalization induced by either phonon environment has complicated time dependence. A thorough investigation into this integral will be published in an additional paper \cite{PaperPrep}. We find that
	\begin{equation}
	\Re\left[I_{ j j}^{\pm}(\omega)\right]=\pi\zeta_{ j}^{\pm}(\mp\omega)F_{ j j}^{\pm}(\mp\omega),
	\end{equation}
	which is given in the main text (\ref{eq:ElecRateE}). To reach an analytic solution for $\zeta_j^{\pm}(\mp\omega)$, we can approximate the continuum of vibrational modes, defined by the spectral density, as a single mode. Specifically, $J(\omega)=\sum_k\left|g_k\right|^2\delta(\omega-\omega_k)\to \left|\tilde{g}\right|^2\delta(\omega-\tilde{\omega})$, where $\tilde{g}$ and $\tilde{\omega}$ are the single mode representation coupling strength and frequency. Estimations for these parameters can be found from the moments of the distribution, $\int_0^{\infty}\rmd\omega\left(J(\omega)/\omega^2\right)\omega^n\equiv m^{(n)}$. One can identify that $m^{(0)}=\phi(0)$ and $m^{(1)}=\lambda$ which are the reorganization energy (\ref{eq:renorm}) and phonon propagator (\ref{eq:PhononProp}) respectively. We can then show that $\left|\tilde{g}\right|^2=m^{(0)}=\phi(0)$ and $\tilde{\omega}=m^{(1)}/m^{(0)}=\lambda/\phi(0)$. Using this, we find that the phonon renormalization at zero temperature is
	\begin{equation}
	\zeta_{ j}^{\pm}(\omega)\approx\kappa_{ j}^2\left[1+\sum_{n=1}^{\infty}\frac{\left[\phi_{ j}(0)\right]^n}{n!}\frac{F^{\pm}_{ j j}\left(\mp\left[\omega-n\frac{\lambda_{ j}}{\phi_{ j}(0)}\right]\right)}{F_{ j j}^{\pm}(\mp\omega)}\right],\label{eq:zetafunctions}
	\end{equation}
	
\begin{figure}[ht!] 
		\centering
		\includegraphics[width=1\linewidth]{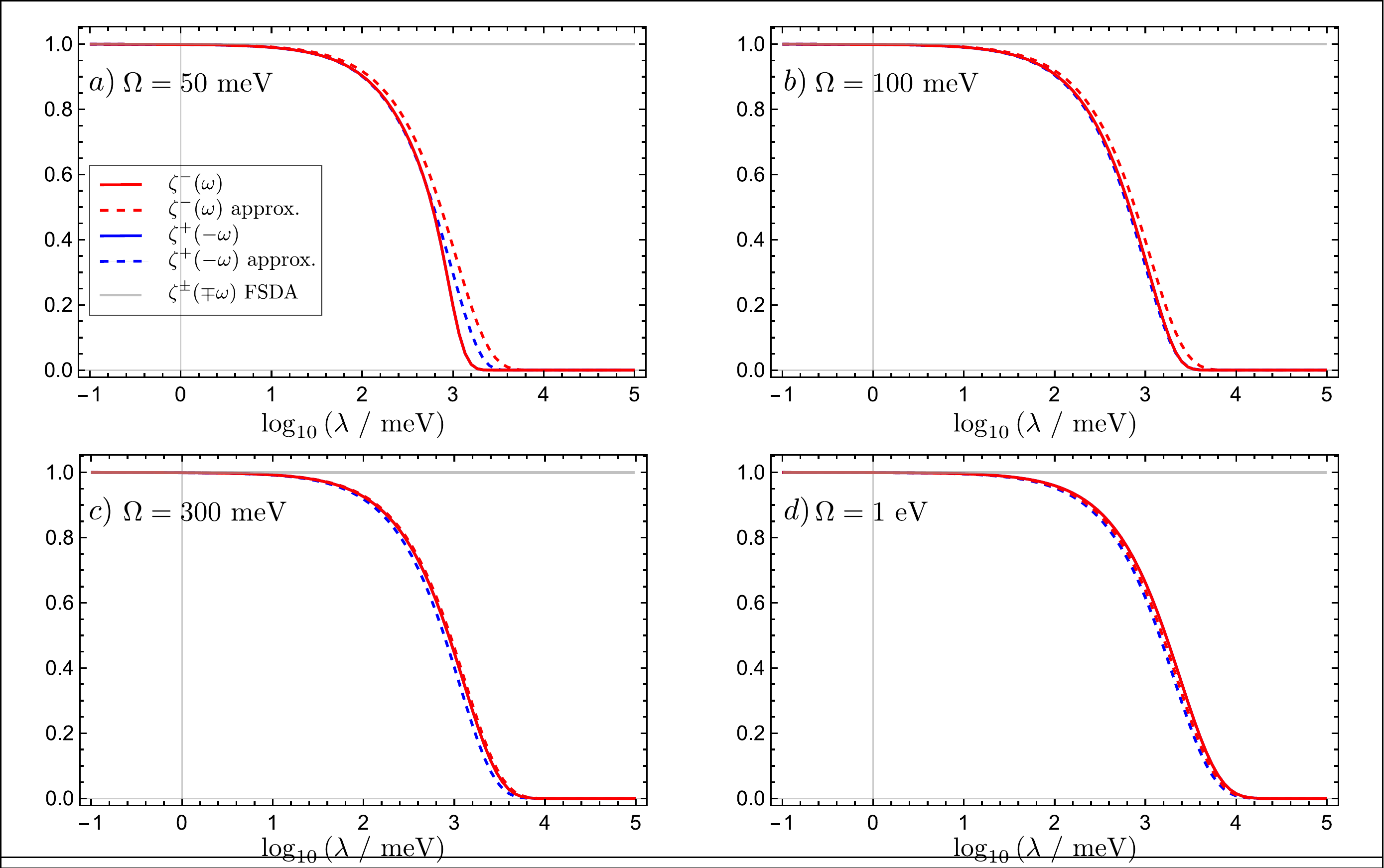} 
		\caption{Comparison of the exact (solid) and approximate (dashed) expressions for $\zeta^{\pm}(\mp\omega)$ (blue and red respectively) plotted against phonon reorganization energy for $J(\omega)=A\omega^3\mathrm{exp}\left(-\omega/\Omega\right)$. Note that the analytic solutions for both $\zeta^{\pm}(\mp\omega)$ are overlapping. We also plot the result from making the FSDA or taking the weak phonon coupling limit (light grey, equal to 1 for all couplings). All curves are generated for zero phonon temperature because this is the condition under which  both the approximate solution is derived. For the example model, we chose a monomer with energy splitting $\omega=2.8$ eV and dipole strength $d=0.2e$ nm. The photon spectral densities with number distributions are $F^-(\omega)=\frac{1}{2\pi}\left[\left(\omega^3d^2\right)/\left(3\pi\right)\right]\left[1+N(\omega)\right]$ and $F^+(\omega)=\frac{1}{2\pi}\left[\left(\omega^3d^2\right)/\left(3\pi\right)\right]N(\omega)$.}
		\label{figA1}
	\end{figure}

	This renormalization can be calculated exactly for certain simple phonon spectral densities. Here, for a single two level system, we test our approximate solution of $\zeta^{\pm}(\omega)$ compared to the exact version for $J(\omega)=A\omega^3\mathrm{exp}\left(-\omega/\Omega\right)$. In Figure~\ref{figA1} we plot the approximated and exact expressions $\zeta^{\pm}(\mp\delta)$ against renormalization energy, $\lambda$ for various cut-off frequencies, $\Omega$. The figure shows reasonable agreement between the two solutions, bearing in mind that under the usual FSDA or weak phonon coupling limit  $\zeta^{\pm}(\mp\omega)=1$ for all reorganization energies. This justifies our using of the approximated form (\ref{eq:zetafunctions}) throughout the paper where we use the spectral density given by (\ref{eq:SD}) for which there is no exact expression for the phonon renormalization. 

	\section{Master equations}\label{App:ME}
	We write, $\dot{\mbox{\boldmath$\rho$}}_{\mathbf{S}}=\left(\mathcal{L}^{\mathrm{S}}+\mathcal{L}^{\gamma}+\mathcal{L}^{\mathrm{pn}}\right)\mbox{\boldmath$\rho$}_{\mathbf{S}}$ where each Liouvillian $\mathcal{L}^x$ is a matrix, and the vector $\mbox{\boldmath$\rho$}_{\mathbf{S}}=\left(\rho_{00},\rho_{++},\rho_{--},\rho_{+-},\rho_{-+} \right)$ where $\rho_{\alpha\beta}\equiv\bra{\alpha}\rho_S\ket{\beta}$. Explicitly, $\mathcal{L}^{\mathrm{S}}\rho_{\mathrm{S}}=-\rmi\left[H_{\mathrm{S}},\rho_{\mathrm{S}} \right]$ and $\mathcal{L}^x\rho_{\mathrm{S}}=\mathcal{D}^x(\rho_{\mathrm{S}})$ for $x=\gamma,\mathrm{pn}$. We ignore the other elements of the density matrix because the dimer level populations decouple from them.

	\subsection{Photon + system Liouvillian}\label{App:PhotonLiouvillian}
	The sum of the system and photon Liouvillians, $\mathcal{L}^{{\mathrm{S}}+\gamma}=\mathcal{L}^{\mathrm{S}}+\mathcal{L}^{\gamma}$ are found as
	\begin{equation} \fl
	\mathcal{L}^{{\mathrm{S}}+\gamma} = \left(
	{\begin{array}{ccccc}
		-[\gamma_+^{\mathrm{A}}+\gamma^{\mathrm{A}}_-]&  \gamma^{\mathrm{E}}_+&  \gamma^{\mathrm{E}}_-& \Theta_+ +\Theta_-^*& \Theta_+^* +\Theta_-\\
		\gamma^{\mathrm{A}}_+&  -\gamma^{\mathrm{E}}_+& 0& -\Theta_-^*& -\Theta_-\\
		\gamma^{\mathrm{A}}_-& 0&  -\gamma^{\mathrm{E}}_-& -\Theta_+& -\Theta_+^*\\
		\tilde{\Theta}_+ +\tilde{\Theta}_-^*& -\Theta_+^*& -\Theta_-&  -\frac{1}{2}[\gamma^{\mathrm{E}}_++\gamma^{\mathrm{E}}_-]-\rmi\tilde{\Delta}& 0\\
		\tilde{\Theta}_+^* +\tilde{\Theta}_-& -\Theta_+& -\Theta_-^*& 0&  -\frac{1}{2}[\gamma^{\mathrm{E}}_++\gamma^{\mathrm{E}}_-]+\rmi\tilde{\Delta}
		\end{array}} \right)
	\end{equation}
	where the absorption rates from the ground state to the $\ket{+}$ or the $\ket{-}$ are denoted $\gamma^{\mathrm{A}}_{\pm}\equiv\Gamma^{\mathrm{A}}_{\pm}N_{\pm}$ and the emission rates from the eigenstates to the ground state are $\gamma^{\mathrm{E}}_{\pm}\equiv\Gamma^{\mathrm{E}}_{\pm}\left(1+N_{\pm}\right)$. The form of $\Gamma^{\mu}_{-}$ is given in the main text for $\mu=\mathrm{A,E}$ (\ref{eq:MinusPhotonRate}), $N_{\pm}$ is the Bose-Einstein distribution evaluated at frequency $\delta_{\pm}$ respectively, and 
	\begin{eqnarray}
	\Gamma^{\mu}_+=\cos^2\frac{\chi}{2}\zeta_1^{\mu}(\delta_+)\gamma_1(\delta_+)&+\sin^2\frac{\chi}{2}\zeta_2^{\mu}(\delta_+)\gamma_2(\delta_+)\\
	&+\kappa_1\kappa_2\sin\chi\sqrt{\gamma_1(\delta_+)\gamma_2(\delta_+)}\mathcal{F}(\delta_+r_{12}).
	\end{eqnarray}
	The renormalized excited state detuning of the dimer is $\tilde{\Delta}\equiv\tilde{\delta}_+-\tilde{\delta}_-$, where
	\begin{eqnarray}
	\fl\tilde{\delta}_+=\delta_++\cos^2\frac{\chi}{2}\Lambda_1(\delta_+)+\sin^2\frac{\chi}{2}\Lambda_2(\delta_+)+\kappa_1\kappa_2\sin\chi\sqrt{\gamma_1(\delta_+)\gamma_2(\delta_+)}\mathcal{G}(\delta_+r_{12}),\label{eq:PlusRenorm}\\
	\fl\tilde{\delta}_-=\delta_-+\sin^2\frac{\chi}{2}\Lambda_1(\delta_-)+\cos^2\frac{\chi}{2}\Lambda_2(\delta_-)-\kappa_1\kappa_2\sin\chi\sqrt{\gamma_1(\delta_-)\gamma_2(\delta_-)}\mathcal{G}(\delta_-r_{12}),\label{eq:MinusRenorm}
	\end{eqnarray}
	and the Lamb shift of the excited state of monomer-$j$ is
	\begin{equation}
	\Lambda_{ j}(\omega)=\frac{\gamma_{ j}(\omega)}{2\pi\omega^3}\mathcal{P}\int_0^{\infty}\rmd zz^3\times\cases{\frac{1+N(z)}{\omega-z},  &for $\omega>0$\\
		\frac{N(z)}{\omega+z}, &for\ $\omega<0$\\} ,
	\end{equation}
	where $\mathcal{P}$ denotes the principal value. This integral diverges but it is assumed that the monomer Lamb shifts are negligible. The coherence generating function (CGF) is defined by the integral
	\begin{eqnarray}
	\fl\mathcal{G}(\omega r_{12})&=\frac{\mathcal{P}}{2\pi \left|\omega\right|^3}\int_0^{\infty}\rmd zz^3\mathcal{F}\left(zr_{12}\right)\times\cases{\frac{1+N(z)}{\omega-z},  &for $\omega>0$\\
		\frac{N(z)}{\omega+z}, &for $\omega<0$\\}\nonumber\\
		\fl&\approx\cases{\mathcal{G}^{\prime}(\omega r_{12})+g(\omega r_{12}), &for $\omega<0$\\
		0, &for $\omega<0$\\}\label{eq:GFunction}
	\end{eqnarray}
	where in the second line we have made the approximation of zero photon temperature. In (\ref{eq:GFunction}) we have defined,
	\begin{equation}
	\mathcal{G}^{\prime}(x)=-\frac{3}{8}\left(\alpha_{12}\frac{\cos x}{x}-\beta_{12}(\frac{\sin x}{x^2}+\frac{\cos x}{x^3})\right),
	\end{equation}
	\begin{eqnarray}
	g(x)=\frac{3}{4\pi}\Bigg[&\alpha_{12}\left(-\frac{1}{x^2}+\frac{1}{x}\left[\sin(x)\mathrm{Ci}(x)-\cos(x)\mathrm{Si}(x)\right]\right)\nonumber\\
	&+\beta_{12}\left(\left[\frac{\cos x}{x^2}-\frac{\sin x}{x^3}\right]\mathrm{Ci}(x)+\left[\frac{\sin x}{x^2}+\frac{\cos x}{x^3}\right]\mathrm{Si}(x) \right)\Bigg],
	\end{eqnarray}
	and we have made use of the standard functions 
	\begin{eqnarray}
	\mathrm{Ci}(x)=-\int_{x}^{\infty}\rmd y\frac{\cos y}{y},\\ 
	\mathrm{Si}(x)=\int_0^{x}\rmd y\frac{\sin y}{y}.
	\label{eq:CosSinInt}
	\end{eqnarray}
	
	Finally, the rates which measure the degree of communication of the excited states with the coherences between them have the forms
	\begin{eqnarray}
	\Theta_{\pm}=\frac{1}{2}\gamma_{+-}^{\mathrm{E}}(\delta_{\pm})\left[1+N_{\pm}\right]+\rmi S_{+-}(\delta_{\pm}),\\
	\tilde{\Theta}_{\pm}=\frac{1}{2}\gamma_{+-}^{\mathrm{A}}(\delta_{\pm})N_{\pm}+\rmi S_{+-}(-\delta_{\pm}),
	\end{eqnarray}
	where we have defined
	\begin{eqnarray}
	\fl\qquad\gamma_{+-}^{\mathrm{E}}(\omega)=&-\frac{1}{2}\sin\chi\left[\zeta_1^{\mathrm{E}}(\omega)\gamma_1(\omega)-\zeta_2^{\mathrm{E}}(\omega)\gamma_2(\omega)\right]\nonumber\\
	&+\kappa_1\kappa_2\cos\chi\sqrt{\gamma_1(\omega)\gamma_2(\omega)}\mathcal{F}(\omega r_{12}),\\
	\fl\qquad\gamma_{+-}^{\mathrm{A}}(\omega)=&-\frac{1}{2}\sin\chi\left[\zeta_1^{\mathrm{A}}(-\omega)\gamma_1(\omega)-\zeta_2^{\mathrm{A}}(-\omega)\gamma_2(\omega)\right]\nonumber\\
	&+\kappa_1\kappa_2\cos\chi\sqrt{\gamma_1(\omega)\gamma_2(\omega)}\mathcal{F}(\omega r_{12}),
	\end{eqnarray}
	and 
	\begin{equation}
	\fl S_{+-}(\omega)=-\frac{1}{2}\sin\chi\left[\Lambda_1(\omega)-\Lambda_2(\omega)\right]+\kappa_1\kappa_2\cos\chi\sqrt{\gamma_1(\omega)\gamma_2(\omega)}\mathcal{G}(\omega r_{12}),
	\end{equation}
	and therefore note that $S_{+-}(-\omega)\approx0$. The Liouvillian shows that due to the imaginary parts of the rates $\Theta_{\pm}$, there are oscillations between the excited state populations and coherences at frequencies $\nu_{\pm}=2\kappa_1\kappa_2\cos\chi\sqrt{\gamma_1(\delta_{\pm})\gamma_2(\delta_{\pm})}\mathcal{G}(\delta_{\pm}r_{12})$. This is the origin of the naming for the CGF. The coherences also have additional oscillations occuring at $\tilde{\Delta}$. If we set $\Theta_+=\Theta_-$ (so that $\nu_+=\nu_-\equiv \nu_{\pm}$), then it can be shown that the combination of these two effects manifests itself in the excited state populations as oscillations with frequency $\nu_{\mathrm{pop}}=\sqrt{\tilde{\Delta}^2+4\nu_{\pm}^2}$. There are corrections to this expression when $\nu_+\neq\nu_-$.
	
	For means of comparison, here we show the Liouvillian for the case where the rotating wave approximation (RWA) is not made in the original photon interaction Hamiltonian. In this case it also becomes necessary to include the double excited state, and only at the very end approximating that it is never populated. This way we can still capture the virtual processes connecting the single excitation eigenstates that pass through the double excited state. We emphasize that we have checked that by making these approximations in the paper we have not lost any information. In this case, the full Liouvillian is
	\begin{eqnarray} \fl
	\mathcal{L}^{\mathrm{S}+\gamma}_f = = \left(
	{\begin{array}{ccc}
		-[\gamma_+^{\mathrm{A}}+\gamma^{\mathrm{A}}_-] & \gamma^{\mathrm{E}}_+ & \gamma^{\mathrm{E}}_-\\
		\gamma^{\mathrm{A}}_+ &  -\gamma^{\mathrm{E}}_+ &  0\\
		\gamma^{\mathrm{A}}_- &  0 &  -\gamma^{\mathrm{E}}_-\\
		\tilde{\Theta}_+ +\tilde{\Theta}_-^* & -\Theta_+^*-\tilde{\Phi}_-^* & -\Theta_--\tilde{\Phi}_+\\
		\tilde{\Theta}_+^* +\tilde{\Theta}_- & -\Theta_+-\tilde{\Phi}_- & -\Theta_-^*-\tilde{\Phi}_+^*\\
		\end{array}}\right.\nonumber
	\\
	\left.
	\qquad\qquad{\begin{array}{cc}
		\Theta_+ +\Theta_-^* & \Theta_+^* +\Theta_-\\
	-\Theta_-^*-\tilde{\Phi}_+^* & -\Theta_- -\tilde{\Phi}_+\\
		-\Theta_+-\tilde{\Phi}_- & -\Theta_+^*-\tilde{\Phi}_-^*\\
		-\frac{1}{2}[\gamma^{\mathrm{E}}_++\gamma^{\mathrm{E}}_-]-\rmi\tilde{\Delta}_f & 0\\
		0 &  -\frac{1}{2}[\gamma^{\mathrm{E}}_++\gamma^{\mathrm{E}}_-]+\rmi\tilde{\Delta}_f\\
		\end{array}}\right)		\label{eq:PhotonLiouvilliannoRWAXX}
	\end{eqnarray}
	where without the RWA the CGF is now given by
	\begin{eqnarray}
	\mathcal{G}(\omega r_{12})&=\frac{\mathcal{P}}{2\pi \left|\omega\right|^3}\int_0^{\infty}\rmd zz^3\mathcal{F}\left(zr_{12}\right)\left[\frac{1+N(z)}{\omega-z}+\frac{N(z)}{\omega+z}\right]\\
	&\approx\cases{\mathcal{G}^{\prime}(\omega r_{12})+g(\omega r_{12})  &for $\omega>0$\\
		\mathcal{G}^{\prime}(\left|\omega\right| r_{12})-g(\left|\omega\right| r_{12}) & for\ $\omega<0$\\} 
	\end{eqnarray}
	where we have again made the approximation of zero photon temperature. Furthermore, the Lamb shifts become
	\begin{eqnarray}
	\Lambda_j(\omega)=\frac{\gamma_j(\omega)}{2\pi \omega^3}\mathcal{P}\int_0^{\infty}\rmd zz^3\left[\frac{1+N(z)}{\omega-z}+\frac{N(z)}{\omega+z}\right],
	\end{eqnarray}
	 but are still assumed to be zero. The additional terms in (\ref{eq:PhotonLiouvilliannoRWAXX}) that originate from virtual processes passing through the double excited state are of the form
	\begin{equation}
	   \fl\tilde{\phi}_{\pm}=\frac{1}{2}\sin\chi\left[\Lambda_1(-\delta_{\pm})-\Lambda_2(-\delta_{\pm})\right]+\kappa_1\kappa_2\cos\chi\sqrt{\gamma_1(-\delta_{\pm})\gamma_2(-\delta_{\pm})}\mathcal{G}(-\delta_{\pm} r_{12}),
	\end{equation}
	and the eigenstate splitting gets further renormalized to $\tilde{\Delta}_f=\tilde{\delta}_{+,f}-\tilde{\delta}_{-,f}$ where
	\begin{eqnarray}
	\fl\tilde{\delta}_{+,f}=\tilde{\delta}_++\sin^2\frac{\chi}{2}\Lambda_1(-\delta_+)+&\cos^2\frac{\chi}{2}\Lambda_2(-\delta_+)\nonumber\\
	\fl&+\kappa_1\kappa_2\sin\chi\sqrt{\gamma_1(-\delta_+)\gamma_2(-\delta_+)}\mathcal{G}(-\delta_+r_{12}),\\
	\fl\tilde{\delta}_{-,f}=\tilde{\delta}_-+\cos^2\frac{\chi}{2}\Lambda_1(-\delta_-)+&\sin^2\frac{\chi}{2}\Lambda_2(-\delta_-)\nonumber\\
	\fl&-\kappa_1\kappa_2\sin\chi\sqrt{\gamma_1(-\delta_-)\gamma_2(-\delta_-)}\mathcal{G}(-\delta_-r_{12}).
	\end{eqnarray}
	\subsection{Phonon Liouvillian}\label{App:PhononLiouvillian}
	We find that
	\begin{eqnarray}
	\fl\mathcal{L}^{\mathrm{pn}} = \left(
	{\begin{array}{ccc}
		0 & 0 & 0\\
		0 &  -\gamma_+^{\mathrm{pn}}(\eta) &  \gamma_+^{\mathrm{pn}}(-\eta)\\
		0 &  \gamma_+^{\mathrm{pn}}(\eta) &  -\gamma_+^{\mathrm{pn}}(-\eta)\\
		0 & 2(iS^{\mathrm{pn}}_{\mathrm{xz}}(0)+\zeta_{\mathrm{xz}}^{*}(\eta)) & 2(iS^{\mathrm{pn}}_{\mathrm{xz}}(0)-\zeta_{\mathrm{xz}}(-\eta))\\
		0 & 2(-iS^{\mathrm{pn}}_{\mathrm{xz}}(0)+\zeta_{\mathrm{xz}}(\eta)) & 2(-iS^{\mathrm{pn}}_{\mathrm{xz}}(0)-\zeta_{\mathrm{xz}}^{*}(-\eta))\\
		\end{array}}\right.\nonumber
	\\
	\left.
	\qquad\qquad{\begin{array}{cc}
		0 & 0\\
		\gamma^{\mathrm{pn}}_{\mathrm{xz}}(0) & \gamma^{\mathrm{pn}}_{\mathrm{xz}}(0)\\
		-\gamma^{\mathrm{pn}}_{\mathrm{xz}}(0) & -\gamma^{\mathrm{pn}}_{\mathrm{xz}}(0)\\
		{-(\bar{\gamma}_+^{\mathrm{pn}}+i\bar{\mu}_+^{\mathrm{pn}})-2\gamma_{\mathrm{zz}}^{\mathrm{pn}}(0)} & (\bar{\gamma}_-^{\mathrm{pn}}-i\bar{\mu}_-^{\mathrm{pn}})\\
		(\bar{\gamma}_-^{\mathrm{pn}}+i\bar{\mu}_-^{\mathrm{pn}}) &  {-(\bar{\gamma}_+^{\mathrm{pn}}-i\bar{\mu}_+^{\mathrm{pn}})-2\gamma_{\mathrm{zz}}^{\mathrm{pn}}(0)}\\
		\end{array}}\right)\label{eq:PhononLiou}
	\end{eqnarray}
	where the rates are given by
	\begin{eqnarray}
	\gamma^{\mathrm{pn}}_{\pm}(\omega)=\gamma_{\mathrm{xx}}^{\mathrm{pn}}(\omega)\pm\gamma_{\mathrm{yy}}^{\mathrm{pn}}(\omega),\\
	\mu_{\pm}^{\mathrm{pn}}(\omega)=S_{\mathrm{xx}}^{\mathrm{pn}}(\omega)\pm S_{\mathrm{yy}}^{\mathrm{pn}}(\omega),\\
	\bar{\gamma}_{\pm}^{\mathrm{pn}}=\frac{1}{2}[\gamma_{\pm}^{\mathrm{pn}}(\eta)+\gamma_{\pm}^{\mathrm{pn}}(-\eta)],\\ \bar{\mu}_{\pm}^{\mathrm{pn}}=\mu_{\pm}^{\mathrm{pn}}(\eta)-\mu_{\pm}^{\mathrm{pn}}(-\eta),
	\end{eqnarray}
	with $\gamma^{\mathrm{pn}}_{ab}(\omega)\equiv2\Re[\zeta_{ab}(\omega)]$ and $S^{\mathrm{pn}}_{ab}(\omega)\equiv\Im[\zeta_{ab}(\omega)]$ for $a,b=x,y,z$. The ECFs are
	\begin{equation}
	\zeta_{ab}(\omega)=\int_0^{\infty}\rmd t\ \rme^{\rmi\omega t}\left\langle B_a^{\dagger}(t)B_b(0)\right\rangle_{\mathrm{E},\mathrm{pn}},
	\end{equation}
	where $B_a(t)$ are given by (\ref{eq:PhononOpx}, \ref{eq:PhononOpy}, \ref{eq:PhononOpz}) for $a=x,y,z$. Using the properties of displacement operators derived in~\cite{nazir2016modelling} we find that 
	\begin{eqnarray}
	\fl\left\langle B_{\mathrm{x}}^{\dagger}(t)B_{\mathrm{x}}(0)\right\rangle_{\mathrm{E},\mathrm{pn}}=\cos^2\chi\frac{C^{\prime 2}}{2}\sinh^2\left[\frac{\phi(t)}{2} \right]=\frac{\cos^2\chi}{\sin^2\chi}\left\langle B_{\mathrm{z}}^{\dagger}(t)B_{\mathrm{z}}(0)\right\rangle_{\mathrm{E},\mathrm{pn}},\\
	\fl\left\langle B_{\mathrm{x}}^{\dagger}(t)B_{\mathrm{z}}(0)\right\rangle_{\mathrm{E},\mathrm{pn}}=\left\langle B_{\mathrm{z}}^{\dagger}(t)B_{\mathrm{x}}(0)\right\rangle_{\mathrm{E},\mathrm{pn}}=\sin\chi\cos\chi\frac{C^{\prime 2}}{2}\sinh^2\left[\frac{\phi(t)}{2} \right],\\
	\fl\left\langle B_{\mathrm{y}}^{\dagger}(t)B_{\mathrm{y}}(0)\right\rangle_{\mathrm{pn}}=\frac{C^{\prime 2}}{4}\sinh\left[\phi(t) \right],
	\end{eqnarray}
	where $\phi(t)$ is defined by (\ref{eq:PhononProp}) and all other combinations equate to zero. The ECF integrals are computed numerically.
	
	The Liouvillian (\ref{eq:PhononLiou}) describes phonon induced excitation from the $\ket{-}$ to the $\ket{+}$ at rate $\gamma_+^{\mathrm{pn}}(-\eta)$ and the reverse process at rate $\gamma_+^{\mathrm{pn}}(\eta)$. It also shows that the eigenstate detuning is increased by $\bar{\mu}_+^{\mathrm{pn}}$. Furthermore, the population eigenstates either diminish or gain through the coherences due to the $\gamma_{\mathrm{xz}}^{\mathrm{pn}}(0)$ term. The sign of this rate determines which population gains and which diminishes. However, unlike the photon interaction, the phonon interaction induces oscillations between the coherences at frequency $\bar{\mu}_-^{\mathrm{pn}}$. Finally, the phonon interaction provides additional decay of the coherences at rate $\bar{\gamma}_+^{\mathrm{pn}}+2\gamma_{\mathrm{zz}}^{\mathrm{pn}}(0)$.

	\section{Approximate analytic dark state condition}\label{App:AnalyticMin}
	Minimizing the rate coefficient (\ref{eq:MinusPhotonRate}) with respect to $z$ allows us to derive the critical dipole ratio for which the destructive interference maximally suppresses the emission rate of the $\ket{-}$. This has previously been calculated for weak vibrational coupling, for homogeneous \cite{creatore2013efficient} and heterogeneous \cite{fruchtman2016photocell} dimers. In both of these cases, however, it was also assumed that the coupling between the monomers $C$ was independent of dipole strength. In our calculations, $C\propto z$ and this means that minimizing (\ref{eq:MinusPhotonRate}) with respect to $z$ can only be done numerically (as we show in Figure~\ref{fig2} of the main text). To derive simple analytic formulae for the location of the dark state and its efficacy once formed we approximate $C^{\prime}$ to be independent of $z$. Really, though, this is just a necessity to find analytic formulae. Despite this, the formulae we derive are largely obeyed when the correct dependence on $z$ is reinstated. To do so, we write $\gamma_2(\omega)=z^2\gamma_1(\omega)$ in (\ref{eq:MinusPhotonRate}) and find the minimum of the rate with respect to $z$. We find that this occurs when $z$ is equal to
	\begin{equation}
	z_{\mathrm{c}}=\tan\frac{\chi}{2}\frac{\kappa_1\kappa_2}{\zeta^{\mathrm{E}}_2(\delta_-)}\mathcal{F}.
	\label{eq:CriticalDipole}
	\end{equation} 
	In this approximation, when a dimer is formed with dipole magnitude ratio $z_c$ the $\ket{-}$ emission rate will be minimal, with the rate coefficient equal to 
	\begin{equation}
	\Gamma_{\mathrm{c}-}^{\mathrm{E}}=\sin^2\frac{\chi}{2}\left[\zeta^{\mathrm{E}}_1(\delta_-)-\frac{1}{\zeta^{\mathrm{E}}_2(\delta_-)}\left(\kappa_1\kappa_2\mathcal{F}\right)^2 \right]\gamma_1(\delta_-).
	\label{eq:CoefficientDark}
	\end{equation}
	Evidently in this simple case, maximal optical decoupling of the minus state occurs for weaker phonon coupling and geometries for which $\mathcal{F}$ is closer to $1$. From (\ref{eq:CriticalDipole}) this also means that $z_{\mathrm{c}}\to 1$, leading to homodimers. If the geometry deviates from $\mathcal{F}= 1$, or there is some degree of phonon coupling, it is not possible for the dimer to become fully decoupled by utilizing destructive interference. We note that under the same conditions proposed in \cite{fruchtman2016photocell}  (that is for very weak phonon coupling and a geometry giving $\mathcal{F}=1$) we recover the same condition on $z_{\mathrm{c}}$ and the resulting $\Gamma_{\mathrm{c}-}^{\mathrm{E}}$ for perfect dark state formation. 
	
	When the dipole-dipole coupling $C$ takes its $z$-dependent form (\ref{eq:DipoleDipoleCoupling}), the qualitative dependencies of the critical dipole ratio $z_{\mathrm{c}}$ (\ref{eq:CriticalDipole}) and rate coefficient $\Gamma_{\mathrm{c}-}^{\mathrm{E}}$ (\ref{eq:CoefficientDark}) on system parameters are still followed when the dark state forms by optimizing destructive interference. This can be seen by comparing these expressions to Figure~\ref{fig2}. For example, we still find that maximal decoupling occurs when $\mathcal{F}=1$ and for weak phonon couplings. However, there are a couple of important consequences of using the full expression of $C$: (a) a minimum with respect to $z$ does not necessarily exist in the rate for a given set of parameters, forming only for fairly homogeneous dimers; and (b) the Bose-Einstein distributions $N(\delta_-)$ now have $z$ dependence, however, we have found numerically that the $z_{\mathrm{c}}$ of the eigenstates have negligible dependence on photon temperature.

	\section{Absorption and emission spectra}
	\subsection{Heterogeneous dimer spectra}\label{App:AsymDimer}
\begin{figure}[ht!] 
		\centering
		\includegraphics[width=1\linewidth]{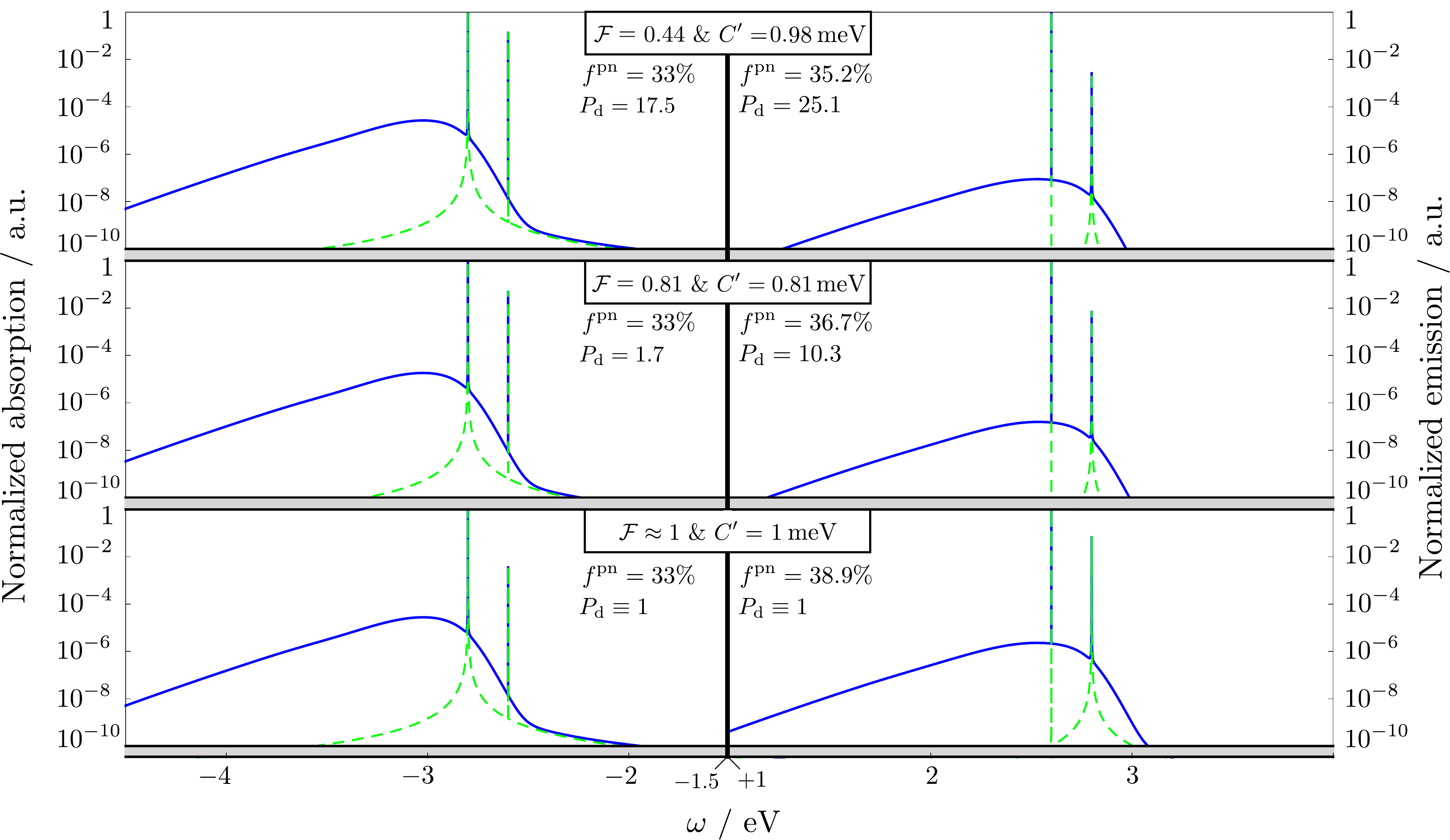} 
		\caption{Absorption and emission spectra for the heterodimer in the same orientations used in Figure~\ref{fig3}. The parameters used in all spectra mean that this dimer corresponds to the HE1 power maxima in Figure~\ref{fig4} but with dipoles in different orientations. The reorganization energy is $\lambda=100$~meV, the renormalized dipole-dipole coupling is $C^{\prime}/z=100$~meV with $z=0.01$ and the renormalized monomer detuning is $\Delta=0.2$~eV. The orientation is varied by row, indicated by the values of $\mathcal{F}$ and $C^{\prime}$. The orientations are the same as those in the homodimer spectra, Figure~\ref{fig3}, and are detailed there. All other parameters are as in previous figures.} 
		\label{figD1}
	\end{figure}

	In Figure \ref{figD1} we plot absorption and emission spectra for the heterodimer, specifically for parameters at the HE1 power maximum in Figure \ref{fig4}. So we can compare with the spectra in the main paper for the homodimer (Figure \ref{fig3}) we choose the same geometries in each row, indicated by $\mathcal{F}$, and the same phonon coupling strength and dipole-dipole coupling strength, with $\lambda=100$ meV and $C^{\prime}=\left(100\ \mathrm{meV}\right)z$. Compared to the homodimer, emission in the heterodimer has much smaller dependence on orientation. This is because the heterodimer is not reliant on destructive interference to create a dark state, and therefore is not sensitive to perturbations from ideal geometries. However, the destructive interference is still playing a role in reducing emission and absorption directly into the dark state which can be seen by the changing value of $P_{\mathrm{d}}$ with orientation.

	\subsection{Calculating spectra}\label{App:Spectra}
	The emission and absorption spectra are calculated as the Fourier transforms of the two-time correlation functions between the positive and negative frequency components of the electric field. It is well documented that, in the Markov approximation, the absorption ($\mu=\mathrm{A}$) and emission ($\mu=\mathrm{E}$) spectrum are given by
	\begin{equation}
	S^{\mu}(\omega)=\Re\int_0^{\infty}\rmd\tau\rme^{\rmi\omega \tau}\lim_{t\to\infty}\sum_{ij=1}^2\gamma_{ij}g_{ij}^{\mu}(t,t+\tau),
	\label{eq:GeneralSpectrum}
	\end{equation}
	where $\gamma_{ij}$ are the (unnormalized) monomer photon rates and $g^{\mu}_{ij}(t,t+\tau)$ are the two-time expectation functions discussed shortly \cite{ficek2005quantum,steck2017quantum}. In (\ref{eq:GeneralSpectrum}), the sum runs through $i,j=1,2$ for the monomers in the dimer. Introducing the notation $\gamma_i(\delta_i^{\prime})\equiv\gamma_i$, the photon rates are given by
	\begin{equation}
	\gamma_{ij}=\cases{\gamma_j  &for $i=j$\\
		\sqrt{\gamma_i\gamma_j}\mathcal{F} & for\ $i\neq j$\\},
	\label{eq:gijrates}
	\end{equation}
	which are the unnormalized single monomer photon rate ($i=j$) and collective rates ($i\neq j$) derived in \ref{App:DarkEmission}. The two-time expectation functions for absorption and emission are
	\begin{equation}
	g_{ij}^{\mu}=\cases{\left\langle\sigma_j^-(t+\tau)\sigma_j^+(t)\right\rangle_{\mathrm{S+E}} & for $\mu=$ A\\
		\left\langle \sigma_j^+(t)\sigma_j^-(t+\tau)\right\rangle_{\mathrm{S+E}}&for $\mu=$ E\\},
	\label{eq:LabFrameTTE}
	\end{equation}
	where $\sigma^{\pm}(t)$ are the raising and lowering operators for monomer-$j$ in the Heisenberg picture \cite{ficek2005quantum,steck2017quantum}. Note that to get the overall absorption of some weak probe-field one must subtract the emission spectrum from the absorption \cite{ficek2005quantum,steck2017quantum}. However, here we are only interested in the raw absorption spectrum. To account for the strong phonon interaction we again move to the polaron frame using the transformation described in Section~\ref{Sec2.2}. It can be shown that after applying the polaron transform, the Heisenberg picture raising and lowering operators become
	\begin{eqnarray}
	\sigma_j^{\pm}(s)\to B_j^{\pm}(s)\sigma_j^{\pm}(s),
	\end{eqnarray}
	where $B_j^{\pm}(s)$ are the displacement operators (\ref{eq:DisplaceOper}) in the Heisenberg picture. We then state the following two-time expectation values of the displacement operators in the Heisenberg picture (for a similar proof see \cite{nazir2016modelling}),
	\numparts
	\begin{eqnarray}
	\left\langle B_i^{\pm}(t+\tau)B_j^{\mp}(t) \right\rangle_{\mathrm{E,pn}}=\cases{\kappa_j^2\rme^{\phi_j(\tau)}& for $i=j$\\
		\kappa_i\kappa_j& for $i\neq j$\\},\\
	\left\langle B_i^{\pm}(t)B_j^{\mp}(t+\tau) \right\rangle_{\mathrm{E,pn}}=\cases{\kappa_j^2\rme^{\overline{\phi}_j(\tau)}& for $i=j$\\
		\kappa_i\kappa_j& for $i\neq j$\\},
	\end{eqnarray}
	\endnumparts
	where $\kappa_j$ and $\phi_j(s)$ are defined in (\ref{eq:kappa}) and (\ref{eq:PhononProp}) respectively, and 
	\begin{equation}
	\overline{\phi}_j(s)=\int_0^{\infty}\rmd \omega\frac{J_{j}(\omega)}{\omega^2}\left[\cos(\omega s)\coth\left(\frac{\beta_{\mathrm{pn}}\omega}{2}\right)+{\rmi}\sin(\omega s)\right],
	\end{equation}
	is the conjugate phonon propagator. Therefore, in the polaron frame, the two-time expectation functions (\ref{eq:LabFrameTTE}) are
	\begin{equation}
	g_{jj}^{\mu}=\kappa_j^2\cases{\left\langle\sigma_j^-(t+\tau)\sigma_j^+(t)\right\rangle_{\mathrm{S}}\rme^{\phi_j(\tau)} & for $\mu=$ A\\
		\left\langle \sigma_j^+(t)\sigma_j^-(t+\tau)\right\rangle_{\mathrm{S}}\rme^{\overline{\phi}_j(\tau)}&for $\mu=$ E\\},
	\label{eq:PolaronFrameTTEjj}
	\end{equation}
	and 
	\begin{equation}
	g_{ij}^{\mu}=\kappa_i\kappa_j\cases{\left\langle\sigma_i^-(t+\tau)\sigma_j^+(t)\right\rangle_{\mathrm{S}} & for $\mu=$ A\\
		\left\langle \sigma_i^+(t)\sigma_j^-(t+\tau)\right\rangle_{\mathrm{S}}&for $\mu=$ E\\}.
	\label{eq:PolaronFrameTTEij}
	\end{equation}
	The two-time expectation values of the lowering and raising operators are evaluated with the quantum regression theorem~\cite{nazir2016modelling,ficek2005quantum,steck2017quantum,mccutcheon2016optical}. This maps these expectation values onto elements of the system density matrix. Therefore, these encode the information on the system dynamics (photon/phonon processes). The phonon sidebands in the spectra arise entirely because of the factors of $\rme^{\phi_j(s)}$ and $\rme^{\overline{\phi}_j(s)}$ in (\ref{eq:PolaronFrameTTEjj}). So, to produce the spectra in Figure~\ref{fig3} and Figure~\ref{figD1} that do not have phonon sidebands (whilst keeping vibrationally induced transfer processes) we have simply set these factors to equal unity.
	\subsection{Approximating $f^{\mathrm{pn}}$}\label{App:fpn}
	The area under the total spectrum is
	\begin{equation}
	A_{\mathrm{T}}^{\mu}=\int_0^{\infty}\rmd\omega S^{\mu}(\omega),\label{eq:AreaSpec}
	\end{equation}
	where $S^{\mu}(\omega)$ is defined in (\ref{eq:GeneralSpectrum}). We can immediately evaluate the integral over frequency using the identity $\Re\int_0^{\infty}\rmd\omega\rme^{\rmi\omega\tau}=\pi\delta(0)$. Then, the integral over $\tau$ in (\ref{eq:GeneralSpectrum}) results in the area being given by the integrand of $\pi S^{\mu}(\omega)$ evaluated at $\tau=0$. The fractional emission or absorption from the phonon sideband is defined as
	\begin{equation}
	f^{\mathrm{pn}}_{\mu}\equiv\frac{A_{\mathrm{T}}^{\mu}-A_{\gamma}^{\mu}}{A_{\mathrm{T}}^{\mu}},
	\end{equation}
	where $A_{\gamma}^{\mu}$ is the integral (\ref{eq:AreaSpec}) with the the factors of $\rme^{\phi_j(s)}$ and $\rme^{\overline{\phi}_j(s)}$ set to unity in the spectrum (the part of the spectrum not generated through the phonon sideband). Before we can evaluate this we must briefly introduce the Markovian quantum regression theorem (QRT) which is used to evaluate the two time system correlation functions in (\ref{eq:PolaronFrameTTEjj}) and (\ref{eq:PolaronFrameTTEij}).

	The theory behind the Markovian quantum regression theorem (QRT) is well documented~\cite{nazir2016modelling,ficek2005quantum,mccutcheon2016optical}, so we will just state the relevant results. One finds from using the QRT, that
	\begin{equation}
	\lim_{t\to\infty}\left\langle \sigma_i^+(t)\sigma_j^-(t+\tau)\right\rangle_{\mathrm{S}}=\chi_{j0}^i(\tau),
	\end{equation}
	where $\chi_{j0}^i(\tau)=\bra{j}\chi(\tau)^i\ket{0}$ and $\chi^i(\tau)$ is an effective density matrix described by
	\begin{equation}
	\frac{d}{d\tau}\chi^i(\tau)=\mathcal{L}^{\mathrm{T}}\chi^i(\tau),
	\end{equation}
	for the same total Liouvillian $\mathcal{L}^{\mathrm{T}}$ that describes the evolution of the system density matrix, $\rho_{\mathrm{S}}(t)$ i.e. $\dot{\rho}_{\mathrm{S}}(t)=\mathcal{L}^{\mathrm{T}}\rho_{\mathrm{S}}(t)$. The superscript $i$ refers to the initial conditions of the density matrix which, through the QRT, are found as
	\begin{equation}
	\chi^i(0)=\rho_{\mathrm{S}}(t=\infty)\sigma_i^+.\label{eq:EmissionICEffective}
	\end{equation}
	Similarly one can show that 
	\begin{equation}
	\lim_{t\to\infty}\left\langle\sigma_j^-(t+\tau)\sigma_i^+(t)\right\rangle_{\mathrm{S}}=\Lambda_{j0}^i(\tau),
	\end{equation}
	with initial condition
	\begin{equation}
	\Lambda^i(0)=\sigma_i^+\rho_{\mathrm{S}}(t=\infty).\label{eq:EmissionICEffective2}
	\end{equation}
	
	We can therefore write general expressions for the areas of the spectrum with and without the phonon sidebands, which are
	\numparts
	\begin{eqnarray}
	A_{\mathrm{T}}^{\mu}=G^{\mu}+\kappa_1\kappa_2 F^{\mu},\\A_{\gamma}^{\mu}=\alpha^{\mu} G^{\mu}+\kappa_1\kappa_2 F^{\mu},
	\end{eqnarray}
	\endnumparts
	where 
	\numparts
	\begin{eqnarray}
	G^{\mu}=\sum_{j=1}^2\gamma_jQ_j^{\mu},\\
	F^{\mu}=\gamma_{12}Q_{12}^{\mu},\\
	\alpha^{\mu}=\frac{1}{G^{\mu}}\sum_{j=1}^2\kappa_j^2\gamma_jQ_j^{\mu}.
	\end{eqnarray}
	\endnumparts
	The functions $Q^{\mu}_{j}$ are the effective density matrices evaluated at zero time. Specifically,
	\begin{eqnarray}
	Q_j^{\mu}=\cases{\Lambda^j_{j0}(0)&for $\mu=\mathrm{A}$\\
		\chi^j_{j0}(0)&for $\mu=\mathrm{E}$\\},\\
	Q_{12}^{\mu}=\cases{\Lambda^2_{10}(0)+\Lambda^1_{20}(0)&for $\mu=\mathrm{A}$\\
		\chi^2_{10}(0)+\chi^1_{20}(0)&for $\mu=\mathrm{E}$\label{eq:Q12x}\\}.
	\end{eqnarray}
	For algebraic ease, we set $\kappa_1=\kappa_2\equiv\kappa$ so that $\alpha^{\mu}=\kappa^2$ for $\mu=\mathrm{A},\mathrm{E}$. Therefore, we arrive at the general expression
	\begin{equation}
	f^{\mathrm{pn}}_{\mu}=\frac{1-\kappa^2}{1+\kappa^2\nu^{\mu}},
	\label{eq:Betax}
	\end{equation}
	where $\nu^{\mu}=F^{\mu}/G^{\mu}$ is the ratio of collective and individual monomer electronic rates.
	
	The zero time values of the effective density matrices concerned with absorption, $\Lambda_{j0}^i(0)$, do not depend on whether the dimer is homogeneous or heterogeneous and can be readily found as
	\numparts
	\begin{eqnarray}
	\Lambda_{j0}^j(0)=P_{00},\\
	\Lambda_{10}^2(0)=\Lambda_{20}^1(0)=0,
	\end{eqnarray}
	\endnumparts
	where we have introduced the notation $P_{ab}\equiv\bra{a}\rho_{\mathrm{S}}(t=\infty)\ket{b}$. Therefore, without approximation,
	\begin{equation}
	\nu^{\mathrm{A}}=0,
	\end{equation}
	and we arrive at (\ref{eq:fpnA}) in the main text.
	Calculating the emission fraction is not so simple, because
	\numparts
	\begin{eqnarray}
	\fl\chi^1_{20}(0)=\sin\frac{\chi}{2}\left[\cos\frac{\chi}{2}P_{++}-\sin\frac{\chi}{2}P_{+-} \right]+\cos\frac{\chi}{2}\left[\cos\frac{\chi}{2}P_{-+}-\sin\frac{\chi}{2}P_{--} \right],\\
	\fl\chi^2_{10}(0)=\cos\frac{\chi}{2}\left[\sin\frac{\chi}{2}P_{++}+\cos\frac{\chi}{2}P_{+-} \right]-\sin\frac{\chi}{2}\left[\sin\frac{\chi}{2}P_{-+}+\cos\frac{\chi}{2}P_{--} \right],\\
	\fl\chi^1_{10}(0)=\cos\frac{\chi}{2}\left[\cos\frac{\chi}{2}P_{++}-\sin\frac{\chi}{2}P_{+-} \right]-\sin\frac{\chi}{2}\left[\cos\frac{\chi}{2}P_{-+}-\sin\frac{\chi}{2}P_{--} \right],\\
	\fl\chi^2_{20}(0)=\sin\frac{\chi}{2}\left[\sin\frac{\chi}{2}P_{++}+\cos\frac{\chi}{2}P_{+-} \right]-\cos\frac{\chi}{2}\left[\sin\frac{\chi}{2}P_{-+}+\cos\frac{\chi}{2}P_{--} \right],
	\end{eqnarray}
	\endnumparts
	where the trigonometric functions describe the rotation between the dipole basis and the eigenbasis and the angle $\chi$ is defined by (\ref{eq:chi1}) and (\ref{eq:chi2}). For an exact solution, one can calculate the steady states numerically, however, here we make approximations to get simpler solutions. We will assume that $P_{--}\gg P_{++}$ which is justified if there is a sufficient bias for phonon transfer to move excitations from the $\ket{+}$ to the $\ket{-}$.
	
	We can further simplify by assuming that the dimer is either homogeneous or heterogeneous. If the dimer is homogeneous, then $\Delta\ll C^{\prime}$ and so $\sin\frac{\chi}{2}\approx\cos\frac{\chi}{2}\approx\frac{1}{\sqrt{2}}$. Furthermore, because the eigenstate populations completely decouple from their coherences in the homodimer we can assume that $P_{+-}=P_{-+}=0$. Under these assumptions, $\chi_{j0}^j(0)\approx\frac{1}{2}P_{--}$ and $\chi_{10}^2(0)\approx\chi_{20}^1(0)\approx-\frac{1}{2}P_{--}\mathrm{sign}\left[C\right]$. Consequently,
	\begin{equation}
	\nu^{\mathrm{homo,E}}\approx-\frac{\gamma_{12}\mathrm{sign}\left[C\right]}{\frac{1}{2}\left(\gamma_1+\gamma_2\right)}\approx-\mathcal{F}\mathrm{sign}\left[C\right],
	\label{eq:SymBetaE}
	\end{equation}
	where in the second equality we have made the further approximation that $\gamma_1\approx\gamma_2$ (true for highly homogeneous dimers) to arrive at (\ref{eq:fpnESymmetric}) in the main text.
	
\begin{figure}[ht!] 
		\centering
		\includegraphics[width=0.8\linewidth]{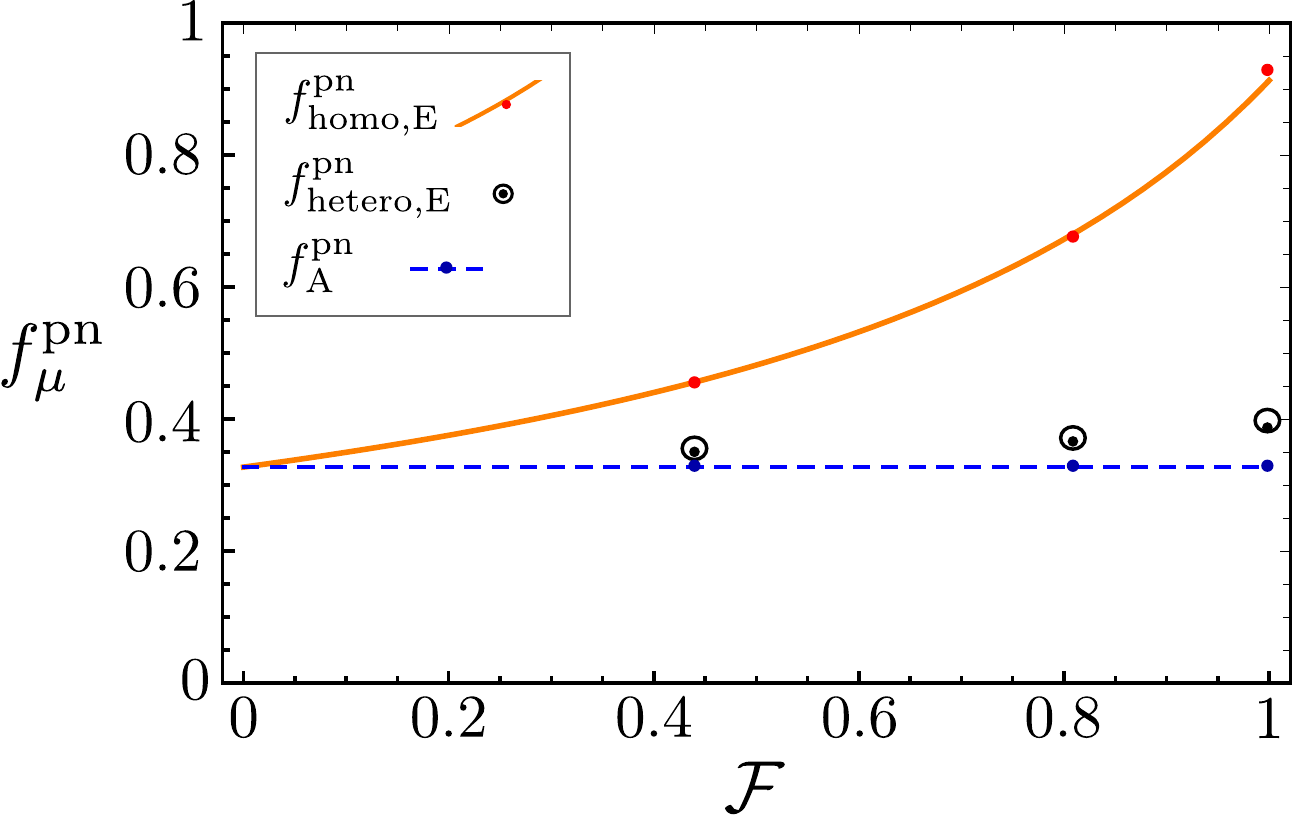} 
		\caption{Comparison between $f^{\mathrm{pn}}_{\mu}$ values calculated from the approximate expressions and and numerically from the homodimer (Figure~\ref{fig3}) and heterodimer (Figure~\ref{figD1}) spectra. The lines and circles are results calculated by substituting the approximate expressions for $\nu^{\mathrm{E}}$ (\ref{eq:SymBetaE}) and (\ref{eq:AsymBetaE}) into (\ref{eq:Betax}). The points are calculated from the spectra. There is overall good agreement. Note that specifying $\mathcal{F}$ does not uniquely determine the orientation of the dipoles. In order to do so one must also specify $C^{\prime}$. Therefore, because $f^{\mathrm{pn}}_{\mathrm{hetero,E}}$ depends on both $\mathcal{F}$ and $C^{\prime}$ we cannot plot a continuous line for the derived expression with respect to $\mathcal{F}$. This is not true for $f^{\mathrm{pn}}_{\mathrm{homo,E}}$ which depends only on $\mathcal{F}$, and not on $C^{\prime}$. This also means that experimentally determining $f^{\mathrm{pn}}_{\mathrm{homo,E}}$ gives direct access to $\mathcal{F}$ and so, for the homodimer, this enables the determination of a set of possible orientations of the dipoles.}
		\label{figD2}
	\end{figure}

	If the dimer is heterogeneous ($\Delta\gg C^{\prime}$), one cannot derive an expression that does not involve the steady states of the density matrix. Defining $\epsilon=C^{\prime}/\Delta\ll 1$, and noting that $\gamma_1\gg\gamma_2$, we find that
	\begin{equation}
	\nu^{\mathrm{hetero,E}}\approx-\gamma_{12}\frac{\epsilon P_{--}-2\Re\left[P_{+-}\right]}{P_{++}\gamma_1+P_{--}\gamma_2},
	\label{eq:AsymBetaE}
	\end{equation}
	where we have used that $\Im\left[P_{+-}\right]\approx 0$. In Figure \ref{figD2} we plot these derived expressions for $f^{\mathrm{pn}}_{\mu}$ with $\mathcal{F}$ for $\lambda=100$~meV. Additionally, we plot the values calculated by numerically integrating the homodimer and heterodimer spectra over all frequencies with and without the phonon sidebands.  

	\section{Idealized load model}\label{App:RC}
	We model the idealized load illustrated in Figure~\ref{fig1}(c) as a two level system called a trap. The evolution of the dimer and trap is described by the total density matrix $\rho(t)\otimes\rho_{\mathrm{t}}(t)$ where $\rho(t)$ and $\rho_{\mathrm{t}}(t)$ are the dimer and trap density matrices. As described in the main text we follow standard Born-Markov procedure and trace out the dimer environment degrees of freedom leaving the relevant ones, $\rho_{\mathrm{rel}}(t)=\rho_{\mathrm{S}}(t)\otimes\rho_{\mathrm{t}}(t)$. Adding the trap with a tensor product was first discussed in~\cite{higgins2017quantum} in place of the original set-up which added the trap energy levels to the system density matrix~\cite{creatore2013efficient,scully2010quantum,dorfman2013photosynthetic}. This was to reduce the number of free parameters needed to describe the extraction process. The evolution of the system density matrix has been derived in the previous sections and is governed by  $\dot{\rho}_{\mathrm{S}}=\mathcal{L}^{\mathrm{T}}\rho_{\mathrm{S}}$ where $\mathcal{L}^{\mathrm{T}}$ is the total Liouvillian. We can write down a phenomenological equation describing the evolution of the trap as $\dot{\rho}_{\mathrm{t}}=-i \big[H_{\mathrm{t}}, \rho_{\mathrm{t}} \big]+\mathcal{D}_{\mathrm{t}}(\rho_{\mathrm{t}})$, where the trap decay dissipator is given by
	\begin{equation}
	\mathcal{D}_{\mathrm{t}}(\rho_{\mathrm{t}})=\gamma_t\big[\ket{\beta}\bra{\alpha}\rho_{\mathrm{t}}\ket{\alpha}\bra{\beta}-\frac{1}{2}\{\ket{\alpha}\bra{\alpha},\rho_{\mathrm{t}} \}\big],
	\end{equation}
	and the trap Hamiltonian is $H_{\mathrm{t}}=\delta_{\alpha}\ket{\alpha}\bra{\alpha}+\delta_{\beta}\ket{\beta}\bra{\beta}$ where $\delta_{\mathrm{t}}=\delta_{\alpha}-\delta_{\beta}$. As described in the main text, there is then a further phenomenological dissipator added which describes non-radiative extraction of excitons from the minus state to the trap. This must act in both spaces and has the form
	\begin{eqnarray}
	\mathcal{D}_{\mathrm{x}}(\rho_{\mathrm{rel}})=\gamma_{\mathrm{x}}\Big[\ket{0}\bra{-}&\rho_{\mathrm{S}}\ket{-}\bra{0}\otimes\ket{\alpha}\bra{\beta}\rho_{\mathrm{t}}\ket{\beta}\bra{\alpha} \\
	&-\frac{1}{2}\{\ket{-}\bra{-} \otimes \ket{\beta}\bra{\beta},\rho_{\mathrm{S}} \otimes \rho_{\mathrm{t}} \}\Big].
	\end{eqnarray}
	Therefore, the evolution of the combined density matrix is described by
	\begin{eqnarray}
	\dot{\rho}_{\mathrm{rel}}&=\dot{\rho}_{\mathrm{S}}\otimes\rho_{\mathrm{t}}+\rho_{\mathrm{S}}\otimes\dot{\rho}_{\mathrm{t}}+\mathcal{D}_x(\rho_{\mathrm{rel}})\nonumber\\
	&=\mathcal{L}^{\mathrm{T}}\rho_{\mathrm{S}}\otimes\rho_{\mathrm{t}}+\rho_{\mathrm{S}}\otimes\bigg(-i \big[H_{\mathrm{t}}, \rho_{\mathrm{t}} \big]+\mathcal{D}_{\mathrm{t}}(\rho_{\mathrm{t}})\bigg)+\mathcal{D}_{\mathrm{x}}(\rho_{\mathrm{rel}}).
	\end{eqnarray}

	\section{Absolute power output}\label{App:AbsolutePower}
\begin{figure}[ht!] 
		\centering
		\includegraphics[width=1\linewidth]{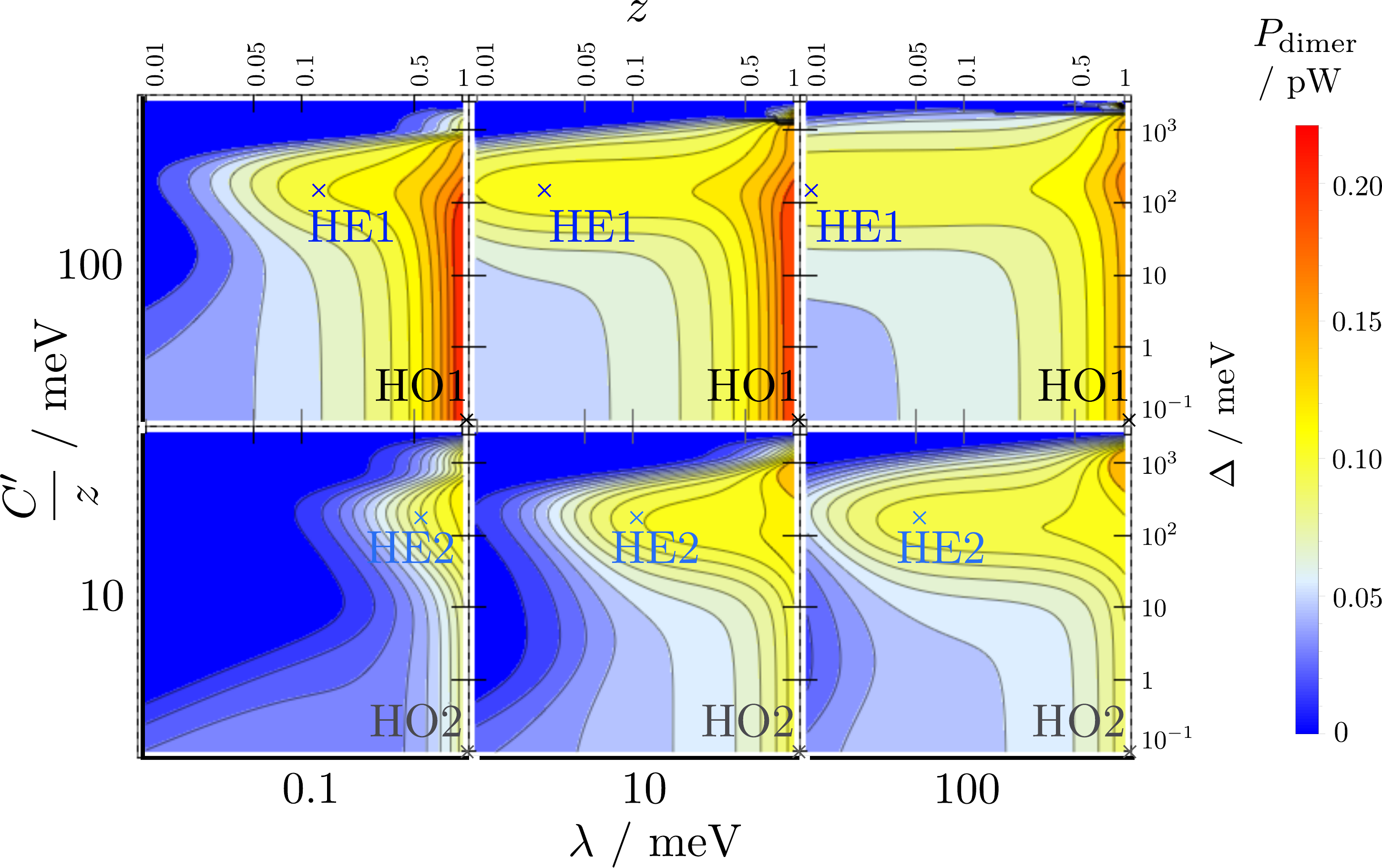} 
		\caption{The absolute power in picowatts of the dimer for the same parameters as used in Figure~\ref{fig4}. We have kept the labels of the homodimer and heterodimer maxima in the same positions on the contour plots.}
		\label{figF1}
	\end{figure}

	Figure~\ref{figF1} shows the power output of the dimer calculated in Figure~\ref{fig4} but without dividing by the power output of the benchmark. The significant difference between the two power plots is that there is now a new power maximum for $z=1$ and large detunings, $\Delta$. This new maximum does not appear in the benchmarked power output because the benchmark similarly maximizes power output here. The reason that the power maximizes here is simply that reducing the energy of monomer-2 will increase the steady state Bose-Einstein population of this excited state. However, there is a trade-off because this decrease in energy of monomer-2 will also mean that the trap voltage has decreased by the same amount. 

	\section{Important rates in determining power output}\label{App:Rates}
\begin{figure}[ht!] 
		\centering
		\includegraphics[width=1\linewidth]{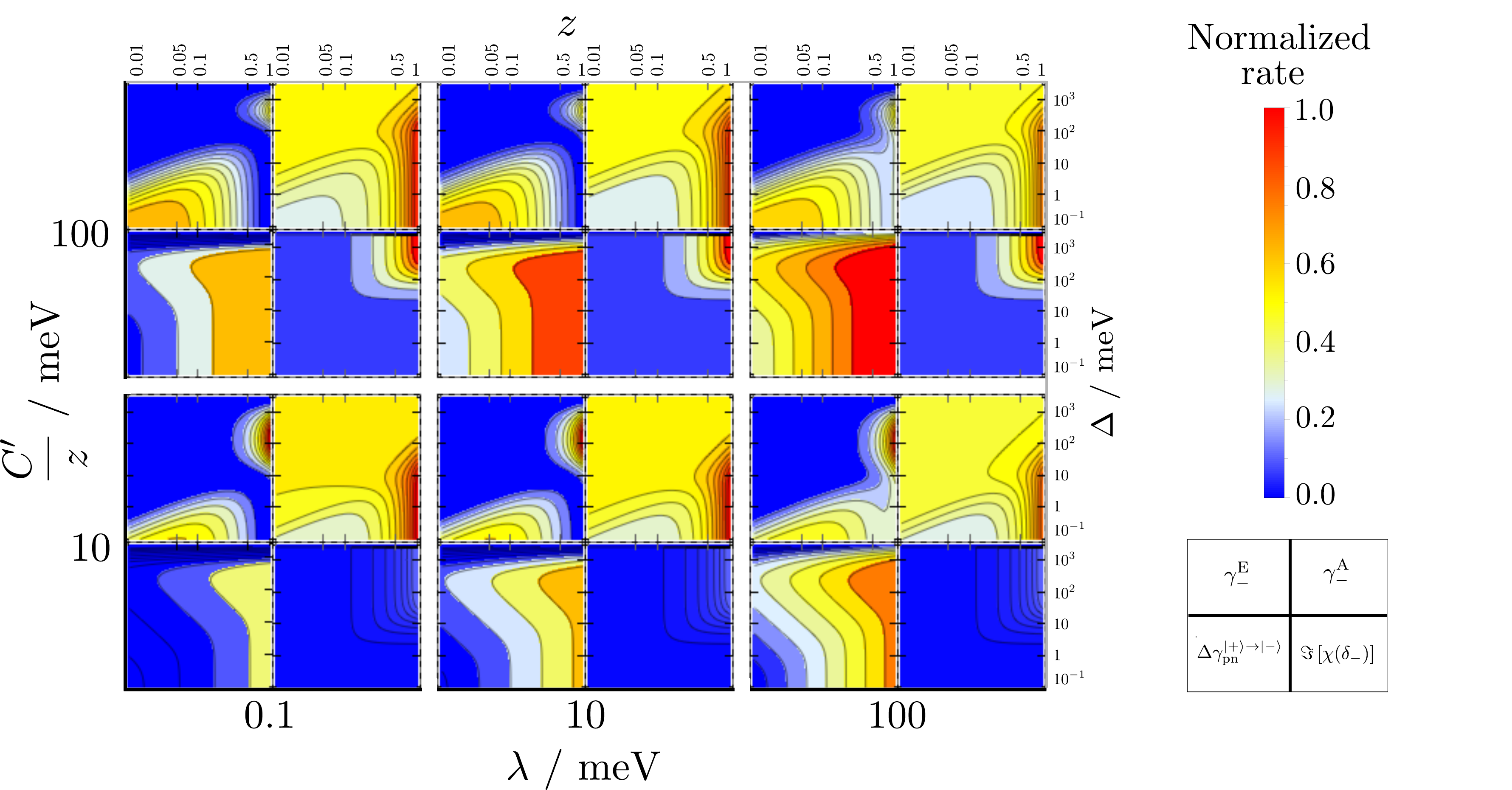} 
		\caption{Contour plots of some important rates in determining the power output. Each set of 4 rate plots correspond to one of the power contour plots in Figure~\ref{fig4} and Figure~\ref{figF1}. Each set shows the $\ket{-}$ emission rate (top left); $\ket{+}$ absorption rate (top right); overall phonon transfer rate from the $\ket{+}$ to $\ket{-}$ (bottom left) and the photon non-secular oscillation frequency from the $\ket{+}$ to the excited state coherences (bottom right). Each rate-type is normalized against the maximum of their type across all six panels. With regard to non-secular oscillation frequency, the effect of decreasing the separation is almost exactly canceled by the effect of increasing the phonon coupling. Note that just like in the power plots, the separations in each panel are different in order to keep $C^{\prime}/z$ fixed. The values are detailed in the caption of Figure~\ref{fig4}}
		\label{figG1}
	\end{figure}

	In Figure~\ref{figG1} we show the important rates in understanding the positions of the maxima in Figure~\ref{fig4} and Figure~\ref{figF1}. Each set of four contour plots in Figure~\ref{figG1} corresponds to the power plot in Figure~\ref{fig4} and Figure~\ref{figF1} which has the same $C^{\prime}/z$ and $\lambda$ values. In each set of four, we show the $\ket{-}$ emission rate (top left); $\ket{+}$ absorption rate (top right); overall phonon transfer rate from the $\ket{+}$ to $\ket{-}$ (bottom left) and the photon non-secular oscillation frequency from the $\ket{+}$ to the excited state coherences (bottom right).
	
	Interestingly, by looking at how the emission rate from the $\ket{-}$ changes as phonon coupling increases for a given dipole-dipole coupling, one can see the dark state in the homodimer, at $(z,\Delta)=(1,0)$, being destroyed. The optical absorption into the $\ket{+}$ is also larger for the HOX than the HEX which prevents bottle-necking and, though less important than a reduced $\ket{-}$ emission, will increase the power output. This occurs because constructive interference of the $\ket{+}$ absorption is equal in magnitude to the destructive interference of the $\ket{-}$ emission. We also show a photon non-secular rate as an example of what these look like.
	\clearpage
	\bibliography{bib}
\end{document}